\documentclass[a4paper,11pt]{article}
\pdfoutput=1 

\usepackage{jinstpub} 
\usepackage{esvect}
\usepackage{subcaption}

\DeclareMathOperator\erf{erf}
\usepackage{siunitx}
\begin{document}
\title{Simulation of charge readout with segmented tiles in nEXO}

\collaboration{\includegraphics[height=17mm]{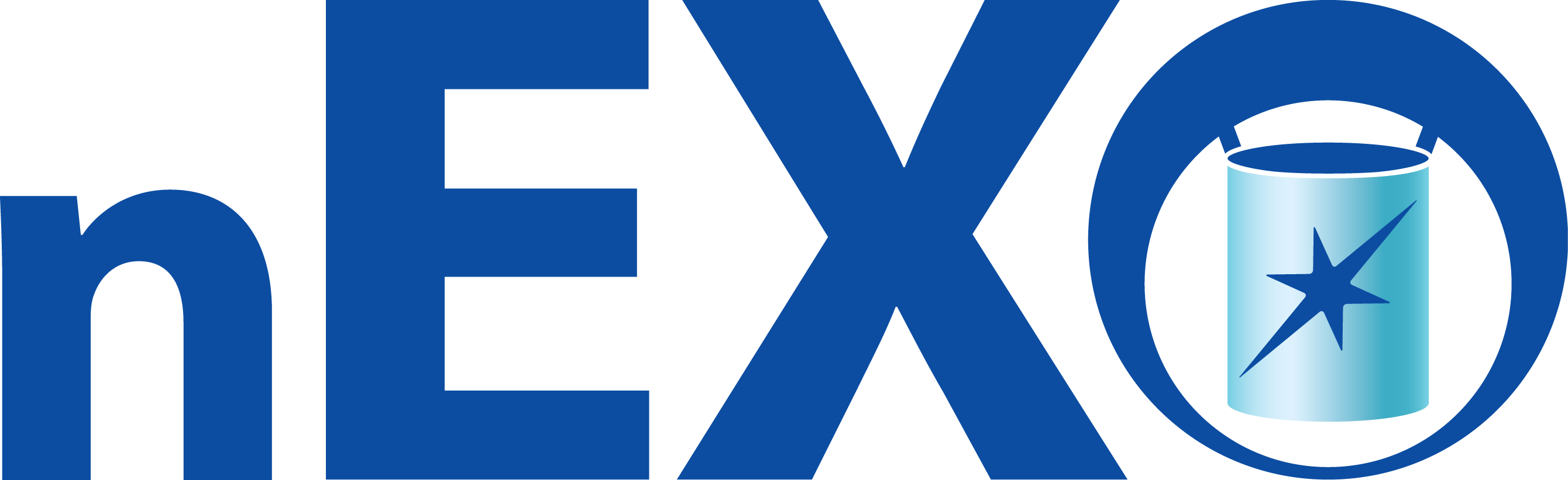}\\ [6pt]nEXO collaboration}
\author[a]{Z.~Li,}
\author[b]{W.R.~Cen,}
\author[c]{A.~Robinson,}
\author[a]{D.C.~Moore,}
\author[b]{L.J.~Wen,}
\author[d]{A.~Odian,}
\author[e]{S.~Al Kharusi,}
\author[f]{G.~Anton,}
\author[g]{I.J.~Arnquist,}
\author[h, 1]{I.~Badhrees,}
\note{also at King Abdulaziz City for Science and Technology, Riyadh, Saudi Arabia}
% ...

\affiliation[a]{Wright Laboratory, Department of Physics, Yale University, New Haven, CT 06511, USA}
\affiliation[b]{Institute of High Energy Physics, Chinese Academy of Sciences, Beijing 100049, China}
\affiliation[c]{Department of Physics, Laurentian University, Sudbury, Ontario P3E 2C6 Canada}
\affiliation[d]{SLAC National Accelerator Laboratory, Menlo Park, CA 94025, USA}
\affiliation[e]{Physics Department, McGill University, Montr\'eal, Qu\'ebec H3A 2T8, Canada}
\affiliation[f]{Erlangen Centre for Astroparticle Physics (ECAP), Friedrich-Alexander University Erlangen-N\"urnberg, Erlangen 91058, Germany}
\affiliation[g]{Pacific Northwest National Laboratory, Richland, WA 99352, USA}
\affiliation[h]{Department of Physics, Carleton University, Ottawa, Ontario K1S 5B6, Canada}
% ...

\abstract{nEXO is a proposed experiment to search for the neutrino-less double beta decay ($0\nu\beta\beta$) of $^{136}$Xe in a tonne-scale liquid xenon time projection chamber (TPC). The nEXO TPC will be equipped with charge collection tiles to form the anode.   In this work, the charge reconstruction performance of this anode design is studied with a dedicated simulation package. A multi-variate method and a deep neural network are developed to distinguish simulated $0\nu\beta\beta$ signals from backgrounds arising from trace levels of natural radioactivity in the detector materials. These simulations indicate that the nEXO TPC with charge-collection tiles shows promising capability to discriminate the $0\nu\beta\beta$ signal from backgrounds. The estimated half-life sensitivity for $0\nu\beta\beta$ decay is improved by $\sim$20$~(32)\%$ with the multi-variate~(deep neural network) methods considered here, relative to the sensitivity estimated in the nEXO pre-conceptual design report.}


\maketitle
\flushbottom
\section{Introduction}
\begin{figure}[htbp]
\begin{center}
\centering
\includegraphics[width=0.8\textwidth]{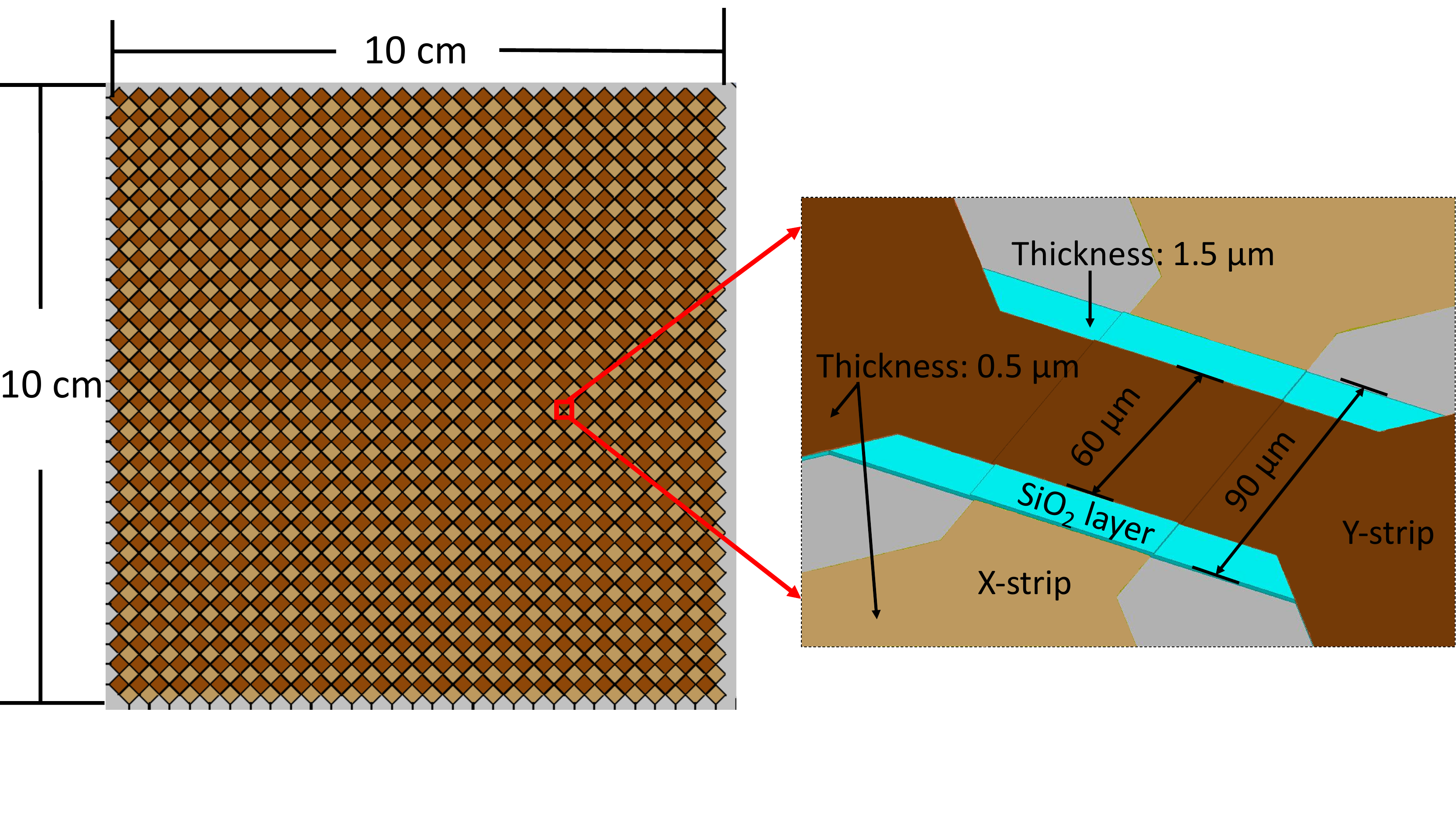}
\caption{Schematic drawing of the prototype charge collection tile described in detail in~\cite{Jewell:2017dzi} with 10 cm edge and 3~mm pitch electrodes (left). In this schematic, the light-colored ``X-strips'' are connected vertically, while the darker ``Y-strips'' are connected horizontally.  Unlike the prototype tile, it is assumed here that the final tile will have a flat edge for the outermost strips so that multiple tiles can be joined to form the anode with no significant gaps in the collection electrodes.  The X/Y traces cross at 60~$\mu$m wide bridges separated by a thin dielectric layer, as shown in a detailed view of the crossing region between strips (right).}
\label{fig:tileprototype}
\end{center}
\end{figure}

\begin{figure}[htbp]
\begin{center}
\includegraphics[width=0.5\textwidth]{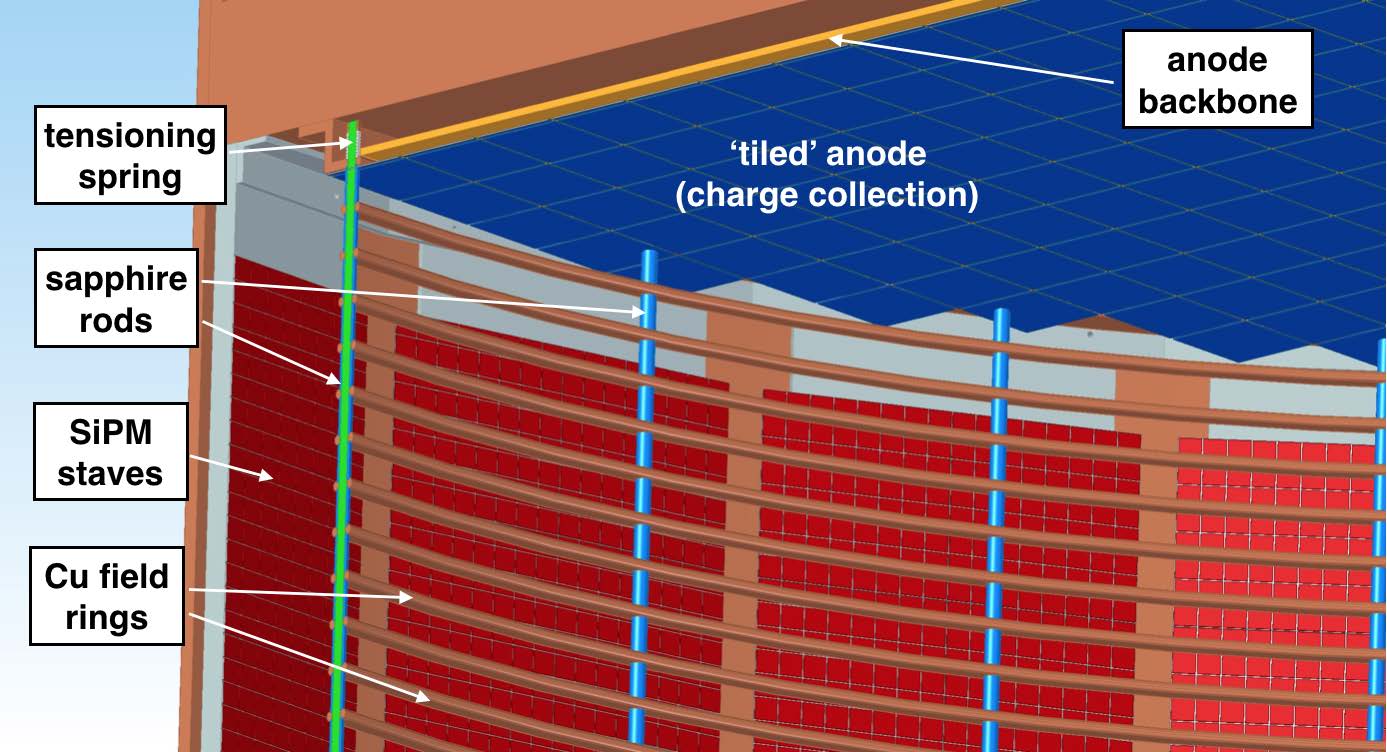}
\caption{A close-up view of the integration scheme of the charge tiles into the nEXO TPC. The anode is composed of a mosaic of many adjacent charge tiles.}
\label{fig:anodemount}
\end{center}
\end{figure}

Neutrino-less double beta decay ($0\nu\beta\beta$) is a hypothetical nuclear decay where two neutrons decay into two protons and two electrons without the emission of anti-neutrinos. The observation of this process would indicate violation of lepton number conservation, and imply that neutrinos are Majorana fermions~\cite{Majorana:1937vz}. 
\par nEXO is a proposed experiment to search for $0\nu\beta\beta$ decay in $^{136}$Xe~\cite{Kharusi:2018eqi}. The nEXO detector would consist of a cylindrical single-phase liquid xenon (LXe) Time Projection Chamber (TPC) filled with 5 tonnes of LXe with 90$\%$ enrichment in $^{136}$Xe. nEXO builds on the success of its predecessor, EXO-200~\cite{exo200_upgraded_0vbb}, which was 36 cm in diameter, and measured
ionization signals with two planes of crossed wires~\cite{Auger:2012gs}. In contrast to EXO-200 and other tonne-scale LXe experiments that deploy meter-long tensioned wire frames as electrodes~\cite{Aprile:2012zx,Mount:2017qzi}, nEXO will implement a segmented anode composed of an array of charge readout tiles. Each tile consists of dielectric substrate covered with an array of conductive strips for charge collection~\cite{Jewell:2017dzi}. A schematic drawing of a 10~cm by 10~cm prototype tile
with a strip pitch of 3~mm is shown in Fig.~\ref{fig:tileprototype}. The tile consists of 60 orthogonal metal charge-collecting strips.  Each strip consists of 30 square pads that are 3~mm across the diagonal and daisy-chained at their corners. Individual strips can be read out as an independent channel. A readout ASIC will be mounted on the reverse side of a tile, and will be connected to the tile via an interposer~\cite{Kharusi:2018eqi}.
Fig.~\ref{fig:anodemount} shows the planned integration scheme of the charge tiles into the nEXO TPC~\cite{Kharusi:2018eqi}. The cathode is located at the bottom of the TPC (not shown), and electrons drift to the anode at the top of the TPC under the influence of the electric field. A prototype charge tile has been tested in a LXe TPC, where good agreement was achieved between the measured ionization spectrum of a $^{207}$Bi source and simulations~\cite{Jewell:2017dzi}.

 Gamma rays from radioactive backgrounds that deposit the same amount of energy as $0\nu\beta\beta$ decays of $^{136}$Xe are the dominant background to the $0\nu\beta\beta$ search. Gamma rays at those energies predominantly have multiple-site interactions due to Compton scattering, while $0\nu\beta\beta$ decays are predominantly single-site. Signal and backgrounds can be distinguished based on the topology of the charge distribution for each event. In this paper, we simulate the charge depositions expected from $0\nu\beta\beta$ decay and backgrounds and the corresponding charge signals produced on charge tiles at the anode. The simulated charge signals are used to construct and optimize event discriminators with machine learning algorithms. The simulation was also used to determine the effect of the charge tile geometry and electric field on the background discrimination for nEXO.

\section{Simulation of ionization electrons in nEXO}
The simulation of the nEXO TPC is split into two stages. The first stage (Sec.~\ref{sec:g4_sim}) uses a GEANT4-based package~\cite{Agostinelli:2002hh, Allison:2006ve} to simulate the production of ionization electrons and scintillation photons by particle interactions in LXe. The second stage (Sec.~\ref{sec:chargesim}) starts from the number and location of ionization electrons produced by the previous package, and simulates the detector signals produced by the drift and collection of electrons on the charge tiles and the response of the electronics.
\subsection{GEANT4-based simulation of energy deposition in nEXO}
\label{sec:g4_sim}
A GEANT4-based package is developed to simulate the energy depositions in the nEXO detector based on a detailed model of the detector geometry. A full list of detector components and physics processes used in the simulation is provided in~\cite{Albert:2017hjq}. Particles interacting with LXe deposit energy by producing both scintillation light (178 nm) and electron-ion pairs (ionization). The production of scintillation and ionization is modeled with the Noble Element Simulation Technique (NEST) package~\cite{NEST2.0}. 
\par In order to analyze the ability of the nEXO TPC to distinguish between $0\nu\beta\beta$ and backgrounds, a collection of $0\nu\beta\beta$ decays and background events are simulated. The dominant backgrounds in the $0\nu\beta\beta$ search arise from $\gamma$s in the $^{238}$U and $^{232}$Th chains that reach the central region of the detector~\cite{Kharusi:2018eqi}. To reduce computation time, only backgrounds arising from the dominant components in the full
detector model are included in the detector Monte Carlo (MC) simulation used in this work, and backgrounds are selected to have their number of ionization electrons within the energy range of interest for $0\nu\beta\beta$ decay events (between 80,000 and 115,000). The number of $^{238}$U and $^{232}$Th events simulated for each of these components is listed in Table \ref{tab:bkgindex}.  Together, the simulated components account for 83$\%$ of the total backgrounds in the full model, providing a good approximation of the dominant backgrounds in the detector.  Of all the simulated events, only events with energy depositions in the innermost 3 tonnes of LXe are used for signal and background discrimination studies, since this region dominates the detector sensitivity~\cite{Kharusi:2018eqi}.

\begin{table}[htp]
\begin{center}
\resizebox{\columnwidth}{!}{%
\begin{tabular}{cccccc}
\hline
Component & Mass (kg)  & \multicolumn{2}{c}{$^{238}$U} & \multicolumn{2}{c}{$^{232}$Th} \\
 & or Area & Total activity (Bq)  & Events simulated &Total activity (Bq) & Events simulated\\
\hline
 HFE-7000 shielding & 31810 & 3.6$\times10^{-3}$ & 2800000 & 1.2$\times 10^{-3}$ & 960000\\
 TPC vessel&447.0& 1.4$\times 10^{-3}$ &1120000 & 2.3$\times 10^{-4}$& 184000\\
 Field ring &68.0 & 2.2$\times 10^{-4}$& 176000& 3.5$\times 10^{-5}$&28000\\
 Support rods and spacers&2.6 & 4.2$\times 10 ^{-4}$&336000 & 6.4$\times 10^{-5}$&51200 \\
 Silicon Photomultiplier (SiPM) cables & 10000 cm$^2$&5.8$\times 10^{-4}$&46400 &5.7$\times 10^{-5}$&45600\\
 SiPM stave&132.4& 4.2$\times 10^{-4}$ & 336000&6.9$\times 10^{-5}$ &55200\\
 SiPM module&11.2& 7.8$\times 10^{-5}$& 62400& 6.9$\times 10 ^{-6}$& 5520\\
 SiPM electronics &2.2 & 3.6$\times 10^{-4}$&288000 & 2.3$\times 10^{-4}$&184000\\
 SiPM solder &0.1& 2.3$\times 10^{-4}$& 184000& 1.75$\times 10^{-4}$& 140000\\
 Charge tile support&34.1& 1.1$\times 10^{-4}$& 88000&1.77$\times 10^{-5}$&14160\\
 Charge tile cables & 2500 cm$^2$ & 1.4$\times 10^{-4}$ & 112000 & 1.4$\times 10^{-5}$ & 11200\\
 Anode solder& 0.1& 2.3$\times 10^{-4}$ & 184000 & 1.75$\times 10^{-4}$&140000\\
 
\hline
\end{tabular}
}
\end{center}
\caption{List of background sources arising from $^{238}$U and $^{232}$Th that are included in the simulation with their respective radioactivity, mass, and the number of primary events simulated. The radioactivities of materials are based on~\cite{Kharusi:2018eqi}.}
\label{tab:bkgindex}
\end{table}%

\subsection{Simulation of electron drift and readout in nEXO}
\label{sec:chargesim}
\begin{figure}[htbp]
\begin{center}
\includegraphics[width=0.95\textwidth]{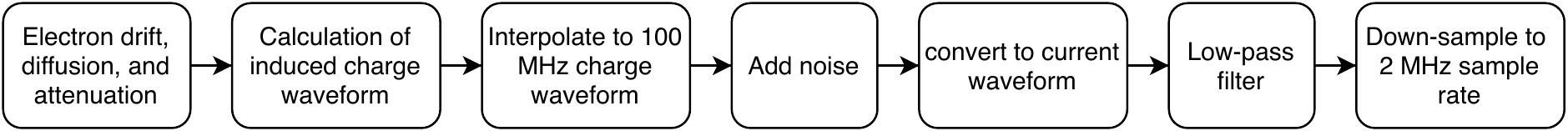}
\caption{Flow chart of the simulation of electron drift and readout in nEXO.}
\label{fig:flowchart}
\end{center}
\end{figure}

The flow of the simulation of electron drift and readout is shown in Fig.~\ref{fig:flowchart}. In the first stage, ionization electrons produced by energy depositions in the LXe are drifted to the anode under the influence of the electric field.  To simulate the detector signals on the anode, each charge deposit is drifted from the interaction location, and the signal induced on each channel is calculated during the drift.  During the drift, electrons can be captured by electronegative impurities, resulting in attenuation of charge signal. In addition, the electrons diffuse transversely and longitudinally. The speed, attenuation, and diffusion of electrons during the simulation of electron drift are modeled based on measurements from EXO-200 and other sources.

\par The electron drift speed in LXe depends on the electric field, and this dependence is modeled based on the measurements in~\cite{PhysRev.166.871}. The drift velocity assumed for the simulations is $\sim$1.9~mm$/\mu$s at an electric field of 380~V/cm, which agrees with recent measurements by the EXO-200 collaboration~\cite{exo200:drift} within 10\%.  As shown in Sec.~\ref{sec:optimization}, the dependence of the results on electric field (which primarily vary due to the change in drift
velocity) does not have a significant effect on background discrimination for velocities between 1.5--2.1~mm/$\mu$s.  The attenuation of ionization electrons in LXe arises from electronegative impurities present in the LXe, which can capture charge as it drifts. The attenuation of the charge signal can be expressed as:
\begin{equation}
\label{eq:qatte}
    N(t)=N_0\exp(-t/\tau)
\end{equation}
where $N(t)$ gives the number of electrons present after drift time, $t$, $N_0$ is the initial number of electrons produced, and $\tau$ is the electron lifetime. 
The nEXO TPC has a length of 125~cm, and nEXO aims to achieve an electron lifetime of $>$10~ms in order to minimize the effects of charge attenuation~\cite{Albert:2017hjq}. Therefore, a lifetime of 10~ms is also assumed in the simulation. In the simulation, the probability of an electron reaching the readout is calculated using its drift length and drift velocity for every ionization electron, following Eq.~\ref{eq:qatte} above. The probability is used to decide whether the electron is saved for further simulation or removed.

\par Electrons diffuse as they drift in LXe, and the diffusion affects classification of 0$\nu\beta\beta$ and background events since it smears out the difference in the initial charge deposit size distribution between the two classes of events. Therefore, it is important to accurately model the diffusion in the simulation.  For $N$ electrons produced at position $\vv{x}_0 = (x_0,y_0,z_0)$ at time $t_0$, the electron distribution at $\vv{x}$ and time $t$ is described by a 3-dimensional diffusion equation:
\begin{equation}
    n(\vv{x},t)=\frac{N}{8 D_T\sqrt{D_L} [\pi (t-t_0)]^{3/2}}\text{exp}\left[\frac{-(x-x_0)^2-(y-y_0)^2}{4D_T (t-t_0)}\right]\text{exp}\left[\frac{-([z-z_0]  - v_d [t-t_0])^2}{4D_L (t-t_0)}\right]
\label{eq:diff}
\end{equation}
where electrons drift in the $z$ direction with a velocity of $v_d$ and $D_T$ $(D_L)$ is the transverse (longitudinal) diffusion coefficient, describing diffusion in the direction perpendicular  (parallel) to the electric field. The electric field dependence of the transverse diffusion coefficient is based on measurements by EXO-200 and previous experiments~\cite{exo200:drift, Doke:1982tb}, and assumes a value of 53~cm$^2/$s at an electric field of 380~V/cm.
The electric field dependence of the longitudinal diffusion coefficient is based on recent measurements within the nEXO collaboration for electric fields ranging between 80~V/cm and 800~V/cm~\cite{sbu:longitudinal}.  The effects of longitudinal diffusion are typically sub-dominant to the transverse diffusion at the fields of interest for nEXO.

The electrons are diffused before drifting to allow the large number of thermal electrons produced by the GEANT4-based simulation to be binned into coarser voxels before the signal generation stage.  The diffused electrons' $z$ position with longitudinal diffusion is converted to drift time with the drift velocity appropriate for the specific electric field, then sampled with cubic voxels with a 3~mm edge in the $x$ and $y$ directions and a length corresponding to a 2~$\mu$s drift time in the $z$ direction. Each voxel of electrons is tracked as it drifts from the production location to the anode assuming a uniform drift velocity along the drift direction. 
\par As each charge voxel drifts to the anode, the induced charge on each electrode is calculated. The charge per unit area induced by a point charge on a conducting plane, $\sigma$, is given by the method of images as:
\begin{equation}
\sigma = \frac{-Q_0 z}{2\pi(x^2+y^2+z^2)^{3/2}}
\end{equation}
where $Q_0$ is the charge, and $x$, $y$ and $z$ are the distances between the point charge and a point on the conducting plane along the $X$-axis, $Y$-axis, and $Z$-axis. The induced charge on a rectangle in the $X$--$Y$ plane that extends from $x_1$ to $x_2$ along the $X$-axis and $y_1$ to $y_2$ along the $Y$-axis can therefore be calculated as:
\begin{equation}
Q(z) = -\frac{Q_0 z}{2\pi}\int_{x_1}^{x_2}\int_{y_1}^{y_2}\frac{dxdy}{(x^2+y^2+z^2)^{3/2}}
\end{equation}
Positive Xe ions and positive holes produced in the ionization process also induce charge on the charge tile. Therefore, a correction is made to account for that charge. Because ions and positive holes in LXe drift with a much smaller velocity than electrons~\cite{hilt1994positive}, they are assumed to be static during the drift of electrons. The charge induced by the Xe ions on the charge tile is calculated as:
\begin{equation}
Q_{ion}=-Q(z_0)
\end{equation}
where $z_0$ is the distance between the initial charge deposit and the charge tile along the direction of the electron drift. Finally, an additional correction is applied to account for the effect of the induced charge on the cathode. This correction is approximated by the mirror charge from the ions induced on an infinite plane (i.e., neglecting edge effects due to the finite extent of the cathode). The correction is expressed as
\begin{equation}
Q(z) = \frac{Q_0 (2L_{max}-z)}{2\pi}\int_{x_1}^{x_2}\int_{y_1}^{y_2}\frac{dxdy}{(x^2+y^2+(2L_{max}-z)^2)^{3/2}}
\end{equation}
where $L_{max}$ is the maximum drift length of electrons in LXe (i.e., the distance between the anode and cathode). The formation of signals on the anode is performed with the assumption of an infinite anode plane. The approximation is valid in the bulk of LXe, but not near the cathode and field rings.  However, since the nEXO sensitivity is dominated by events in the central region of the detector~\cite{Kharusi:2018eqi}, this approximation is sufficient for modeling background discrimination in the inner 3~tonnes.  Future work will allow accurate simulation of the signal shape of events near the edge of the TPC, which are important for fitting the background event energy spectrum.

\par
\begin{figure*}[htbp]
\centering
    \begin{subfigure}[t]{0.45\textwidth}
        \centering
        \includegraphics[width=\textwidth]{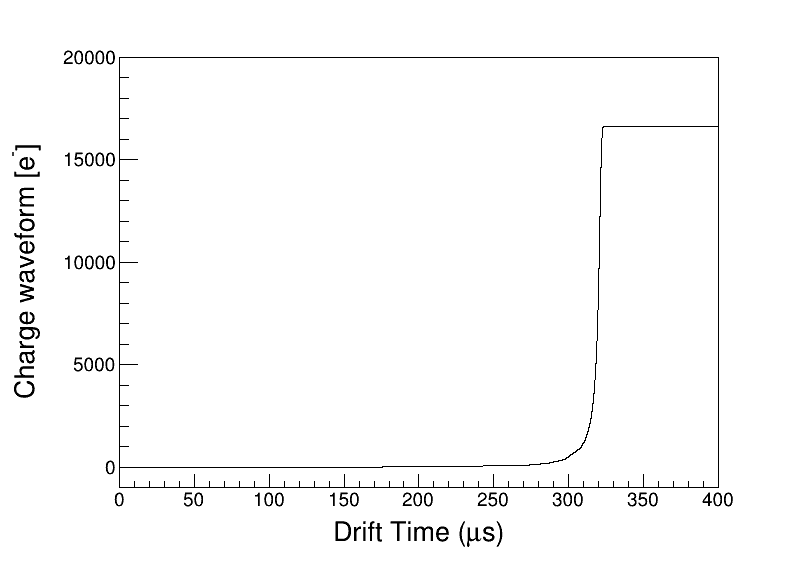}
	\caption{An example of charge signal waveform with 100 MHz sampling rate, prior to adding noise. This channel collects $\sim$16,000 electrons with drift time of $\sim$325~$\mu s$. }
	\label{fig:100MHz}
    \end{subfigure}%
    \hspace{0.09\textwidth}
    \begin{subfigure}[t]{0.45\textwidth}
        \centering
        \includegraphics[width=\textwidth]{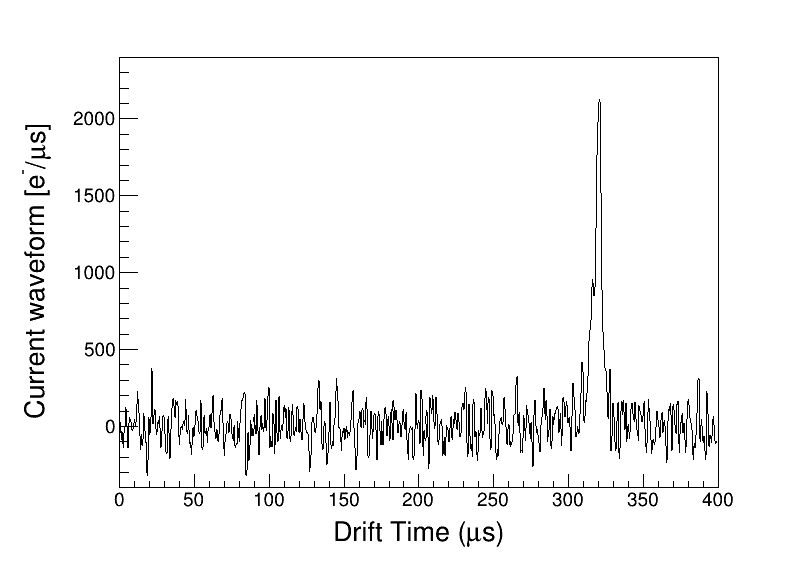}
        \caption{The current signal waveform with 2 MHz sampling rate for the same channel in Fig. \ref{fig:100MHz} after addition of noise. }
\label{fig:wf_2MHz}
    \end{subfigure}
    \caption{An example waveform on a single channel at different stages of the simulation.}
   \end{figure*}

 In order to reduce computing time, the simulation uses an unequal binning of sampling points along the drift direction. When the voxel's distance to the charge tile is smaller than 10~mm, the sampling points are uniformly spaced by 0.5~$\mu$s in time (which is a factor of two smaller than the inverse of the Nyquist frequency for the 2~MHz sampling rate of the electronics, as described below). When the voxels are farther than 10~mm from the charge tile, the sampling interval increases with the
distance, with sampling points located at 10, 15, 20, 40, 60, 80, 100, 300, 500, 700, 900, 1100, 1300, and 1350~mm. The sampling points are selected to both optimize computing time and preserve the information of the waveform. Signals are only generated for pads with charge collection, and for pads within a distance of 9~mm to a charge collection. The sparse sampled waveform is then linearly interpolated to produce a waveform with a uniform 100 MHz sampling rate. Fig. \ref{fig:100MHz} shows an example of a charge signal waveform with 100 MHz sampling rate, prior to adding noise.

\par nEXO plans to utilize in-LXe cold electronics to read out the induced current signals, rather than the integrated charge.  The total charge can then be reconstructed in software by integrating the current waveform. The cold electronics are designed to have an equivalent noise charge (ENC) $<$ 200 e$^-$~\cite{Kharusi:2018eqi} on each anode strip. The simulated noise is first produced in the frequency domain by sampling an assumed noise spectrum for frequencies between 0 and 100 MHz. To determine the impact of
possible variations in the noise spectrum for the final nEXO electronics, a white noise spectrum with constant amplitude between 0 and 100~MHz and a noise spectrum with a peak amplitude at 500~kHz (motivated by measurements of prototype cold electronics~\cite{Radeka:2011zz}) were simulated.  In both cases, the noise amplitude was normalized to the same ENC regardless of spectral shape.  The difference in spectral shape was found to have negligible effect on the results shown in
Sec.~\ref{sec:bdt_discrim} when both were normalized to the same ENC.  Thus, we do not expect that the detailed shape of the noise spectrum has a significant impact on these results. The coherent noise and cross-talk are assumed to be negligible based on tests of a small system in~\cite{Jewell:2017dzi}. Further studies and tests are necessary to understand these issues in nEXO. The waveform of the charge signal in the time domain is transformed to the frequency domain with a fast Fourier
transformation (FFT), and it is then added to the generated noise spectrum. The phase of the noise is randomly generated between 0 and 2$\pi$. The simulated charge noise is scaled to have an RMS of 200 electrons in the time domain after applying the low-pass filter described below, corresponding to the expected ENC above. 

Since nEXO will readout a current waveform rather than the integrated charge, the high-bandwidth charge waveform is converted to a current waveform by calculating its derivative prior to applying the analog anti-aliasing filter that will be used in the nEXO cold electronics. In the real detector, the current waveform will be directly recorded by the electronics.  However, since the derivative and FFT operations relating the charge and current waveforms are linear, it is convenient to perform the previous steps using the simulated charge waveform, and then to convert to a current waveform prior to filtering.  Due to this linearity, the ordering above produces the same current waveform as would be directly recorded by the detector electronics. After adding the noise, a low-pass filter is implemented with a cut-off frequency of 300 kHz.  This filter is applied to the current waveform in the frequency domain, and the waveform is transformed back to the time domain with an inverse FFT. 
The current waveform is then down-sampled to the 2~MHz sampling rate that will be used in the nEXO electronics. This current waveform is saved for each channel with non-zero induced charge, and for the noise channels described below. Fig. \ref{fig:wf_2MHz} shows an example of a simulated waveform with a sampling frequency of 2 MHz.

For a given event, many channels do not have any simulated signal other than noise, since the induced charge is zero.  However, noise present on those channels can still lead to spurious signals when reconstructing data from a real detector.  To account for these channels, while avoiding the need to generate many noise waveforms with no signals above threshold, a single charge value is sampled from a Gaussian distribution with a standard deviation of 200~e$^-$ for each channel. A waveform with only electronics noise is then generated for each channel with a charge arising from
electronics noise greater than 3$\sigma$ (600 e$^-$), and the collected charge on the channel is set to be the randomly generated noise. The drift time of charge signal on a noise channel is randomly assigned to be between 0 and the maximum drift time.  This procedure is expected to produce the same signal and time distribution as would be found from reconstructing signals in a large number of noise traces, but substantially reduces computation time.

\subsubsection{``Fast'' charge simulation}
In addition to the detailed charge simulation described above, a Python application implementing a ``fast'' charge simulation was also developed to simulate the drift, diffusion, and collection of charge deposits in LXe.  The purpose of this application is to avoid the simulation of the microscopic details of the charge deposition and drift in exchange for computational efficiency.  Comparison of the results of the fast charge simulation against the detailed charge simulation (see Sec.~\ref{sec:fast_charge_val}) can validate its use in certain studies of the detector performance, although the detailed simulation is primarily used for all discrimination studies performed here.
\par In order to minimize computation time for the preceding charge simulation, the fast charge simulation drifts electrons converted from energy depositions  in LXe, rather than drifting the resulting thermal electrons. An approximate value of \SI{25}{e\volt} per electron for both $\beta$ (signal) and $\gamma$ (background) events is assumed based on NEST~\cite{NEST2.0}.  This value is approximate and does not include fluctuations expected in a detailed microscopic simulation. The electric field dependence of the drift velocity ($v_{d}$) and diffusion coefficients ($D_{t}$ and $D_{l}$) are the same as that used in the detailed charge simulation. Electrons converted from energy deposits are drifted and diffused in the fast simulation using an approximation of the analytic density distribution $Q$ (a three-dimensional Gaussian) on the anode.  The probability distribution describing the density of the charge distribution is $P(x,y,t) = n(\vv{x},t)/N$ as defined in Eq.~\ref{eq:diff}, so that the charge distribution is given by $Q(x,y,t) = Q_{total}P(x,y,t)$. The fraction of the total charge, $H_V$, collected by a given anode pad is then found by integrating this distribution:
\begin{align}
H_{V} &= \int\limits_{V}dV\,P(x,y,t) \\
&= \left. \frac{1}{8} \erf\left(\frac{x-x_{c}}{\sqrt{4D_{t}t_{c}}}\right)\, \erf\left(\frac{y-y_{c}}{\sqrt{4D_{t}t_{c}}}\right)\, \erf\left(\frac{t - t_{c}}{\sqrt{4(D_{l}/{v_{d}}^{2})t_{c}}}\right)  \right|_{V} \label{eq:timeintegral}
\end{align}
where $dV = dx\,dy\,v_{d}dt$, $V$ is the 3D volume collected by a given pad on an anode strip across a span of time (a sampling interval), $\erf$ is the \emph{Gauss error function}, and $Q_{V} = Q_{total} H_V$ is the charge collected by the element during that time.  The coordinates $x_c$, $y_c$, and $t_c$ are the position and time of charge collection without diffusion. The charge is integrated over a region that extends to at least $3\sigma$ from each deposit on each axis.  
Exponential charge attenuation is further incorporated by shifting the mean and reducing the amplitude of the density function, as can be derived from including the linear exponent in the Gaussian and solving the resulting quadratic equation in time:
\begin{align}
    \exp(-t/\tau)Q_{total}P(x,y,t) &= \left( \exp(-\frac{t_{c} - {\sigma_{l}}^{2}/{2\tau}}{\tau}) Q_{total} \right) P\Big(x,y, \big(t +\tfrac{{\sigma_{l}}^{2}}{\tau} \big) \Big)
\end{align}
where ${\sigma_{l}}^{2} = 2(D_{l}/{v_{d}}^{2})t_{c}$ and $\tau$ is the electron lifetime.
\par The total charge collected on a strip over time for an event is smeared by Gaussian noise, and charge collections exceeding the detection threshold on a given channel are recorded. Unlike the detailed charge simulation, only the total collected charge on each channel is produced by the fast charge simulation, and detector waveforms are not generated.  The collected charge can then be compared against the reconstructed waveform quantities described in the following sections.

\section{Discrimination of $0\nu\beta\beta$ decays and backgrounds based on the reconstruction of ionization electrons}
\subsection{Reconstruction of ionization electrons}
\label{sec:recon_vars}
For the waveforms produced by the detailed charge simulation, each current waveform is first converted to a charge waveform by cumulatively integrating the current waveform, and charge is reconstructed as the average of the last 20 points on the charge waveform.  For the 2~MS/s sampling rate considered here, this corresponds to an average over frequencies $>100$~kHz, for which sources of low-frequency noise in the nEXO detector, which may not be included in the simple model used here, are expected to be small.  Only the charge greater than a threshold is saved to a reconstructed event for later use in constructing discriminators.  Although a low threshold is desired to capture as many signals as possible, at too low of a threshold the rate of false (noise-induced) signals becomes significant, degrading discrimination. To determine the optimal threshold for this cut, a series of threshold values were scanned, and a value of 4.5 times the RMS of channel noise 
was found to be the optimal threshold value for discriminating between signal and backgrounds. While this threshold is used for the multi-variate analysis described below, a threshold-free analysis is also presented in Sec.~\ref{sec:dnn}. 
A ``hit'' channel is also assigned a position as the center of the strip on the anode in the $x$-$y$ coordinate plane. A hit channel that extends on an $x$-axis ($y$-axis) has a precise $y$ ($x$) position and imprecise $x$ ($y$) position of the collected charge in reconstruction. 

\par
\begin{figure}[!ht]
\begin{center}
\includegraphics[width=0.5\textwidth]{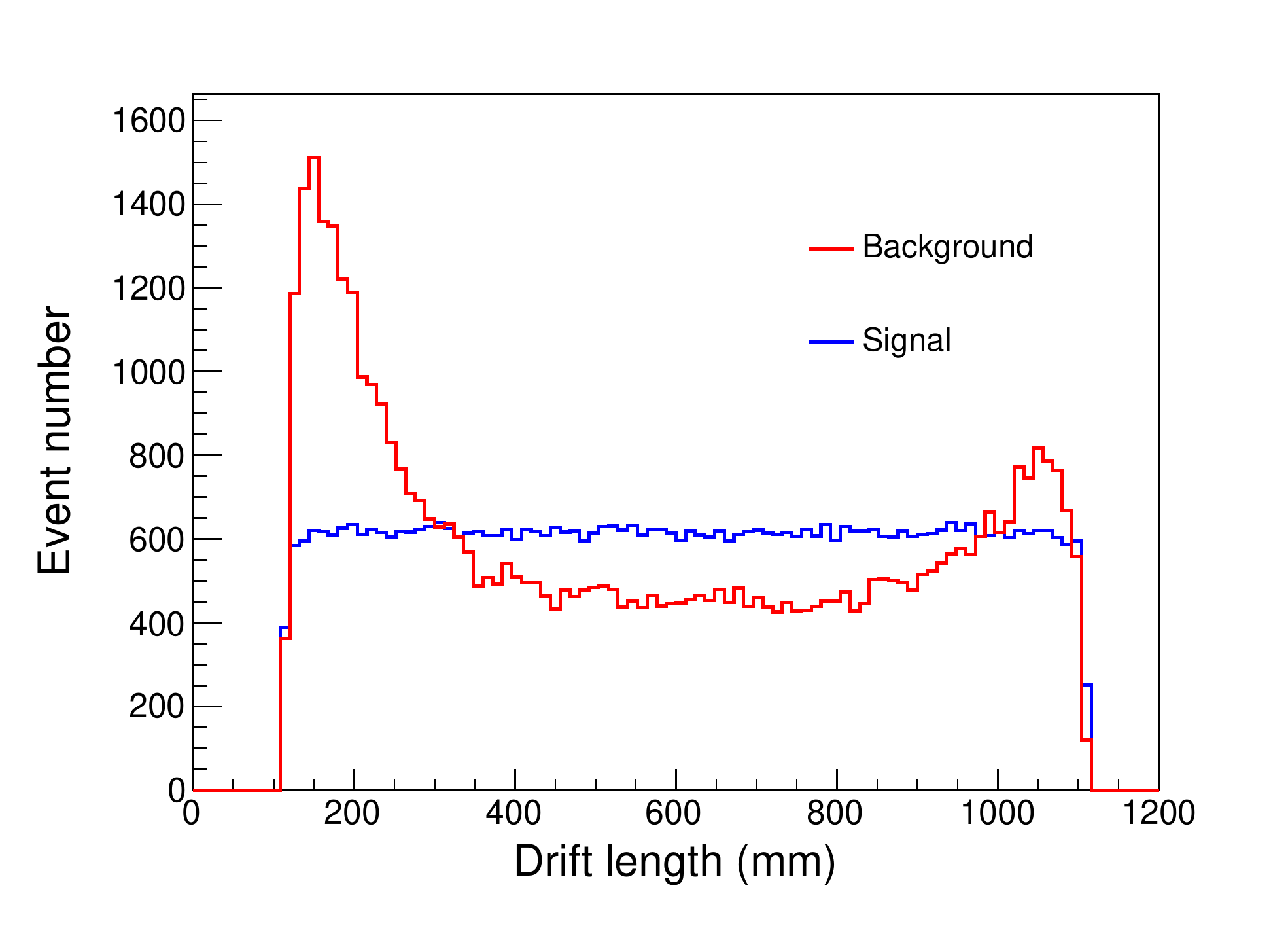}
\caption{Drift length of $0\nu\beta\beta$ decays (blue) and background events (red) in the innermost 3~tonnes of the LXe calculated using the MC truth position of the ionization electrons. The drift length distribution extends only between 100~mm and 1120~mm due to the cut requiring events to lie in the innermost 3~tonnes of the LXe. The distribution of $0\nu\beta\beta$ decays is uniform within the fiducial region of the detector. The background sample has 51,076 events, and the 0$\nu\beta\beta$ sample is normalized to the same statistics. }
\label{fig:drift_3mm}
\end{center}
\end{figure}

 In nEXO, the detection of scintillation photons will provide a precise measurement of the interaction time, $t_0$. Using this $t_0$, the drift length can then be reconstructed using the electron drift velocity and the drift time between $t_0$ and the charge collection time. For this simulation, the scintillation photons are not directly simulated. Therefore a drift length is calculated with the MC truth position of the ionization electrons by averaging their distance to the anode. Fig.~\ref{fig:drift_3mm} shows the drift length distribution reconstructed for $0\nu\beta\beta$ events and background events.

\par
\begin{figure}[!ht]
\begin{center}
\includegraphics[width=0.5\textwidth]{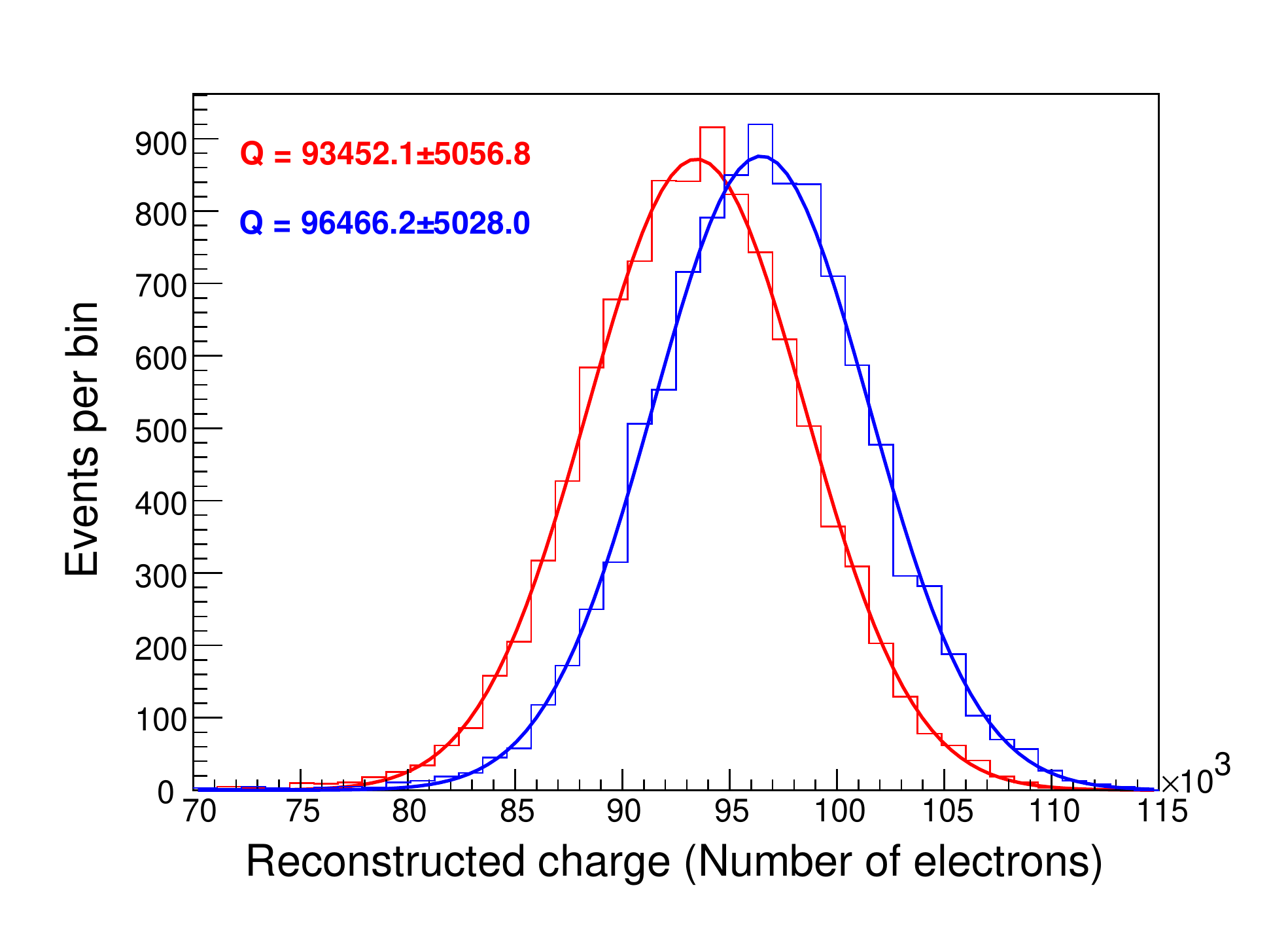}
\caption{Reconstructed charge with (blue) and without (red) the drift time correction in the $0\nu\beta\beta$ simulation. The reconstructed charge is fitted with a Gaussian distribution. }
\label{fig:charge_cor}
\end{center}
\end{figure}

 As described in Sec.~\ref{sec:chargesim}, ionization electrons are attenuated as they drift under the influence of the electric field. This attenuation has a dependence on drift length, so the reconstructed charge of each event is corrected with its drift length based on Eq.~\ref{eq:qatte}. The reconstructed charge before and after the drift time correction is shown in Fig.~\ref{fig:charge_cor}. Since the widths of the distributions are dominated by recombination fluctuations, the correction does not substantially change the charge-only resolution.

\subsection{Reconstruction of event topology}
\label{sec:bdt_discrim}
LXe TPCs have the advantage of both good energy resolution and 3D reconstruction of the event topology, which helps to mitigate background events in the search for 0$\nu\beta\beta$ decay. 
To optimize the separation of signal and backgrounds using the reconstructed 3D topology of the events, a multivariate 
method is constructed that incorporates multiple reconstructed variables.
Specifically, a signal and background discriminator based on a boosted decision tree (BDT) is built via the TMVA software~\cite{Hocker:2007ht}.  The BDT is built with the variables listed below after reconstructing them from the waveforms produced by the detailed charge simulation, and trained to optimally distinguish backgrounds and 0$\nu\beta\beta$ decays. The variables used in the BDT are:
\begin{itemize}
\item[1.] ``Charge-averaged distance to the event center'', $d_x$ and $d_y$  in the $x$ and $y$ coordinates. With the reconstructed charge and position of hit channels, a charge-average center position, $\bar{x}$ ($\bar{y}$), in the $x$ ($y$) coordinate of an event is calculated as:
\begin{equation*}\label{eq:center}
\bar x = \frac{\sum x_i\times q_i}{\sum q_i},  \quad \bar y = \frac{\sum y_j\times q_j}{\sum q_j}
\end{equation*}
where $x_i$ ($y_j$) and $q_i$ ($q_j$) is the position and reconstructed charge of the $i^{th}$ ($j^{th}$) hit channel with a precise $x$ ($y$)  position. With the reconstructed center position, a charge-averaged distance to the event center, $d_x$ ($d_y$), is calculated in the $x$ ($y$) coordinate, respectively, as:

\begin{equation}
d_x = \frac{\sum \mid x_i-\bar x\mid\times q_i}{\sum q_i}, \quad d_y = \frac{\sum \mid y_j-\bar y\mid\times q_j}{\sum q_j}
\label{eq:dist_to_center}
\end{equation}
A 0$\nu\beta\beta$ event consists of two electrons with total energy of 2.46~MeV, which typically produce energy deposits that appear to be a single electron cloud at the nEXO spatial resolution. However, a 2.46~MeV $\gamma$-ray has an attenuation length of 8.5~cm, and most $\gamma$-ray interactions involve Compton scattering that produce multiple energy depositions in the detector. As shown in Fig.~\ref{fig:aved}, $0\nu\beta\beta$ events are expected to have a smaller value for the charge-averaged distance to the event center, with respect to backgrounds.

\begin{figure}[htbp]
\begin{center}
\includegraphics[width=0.5\textwidth]{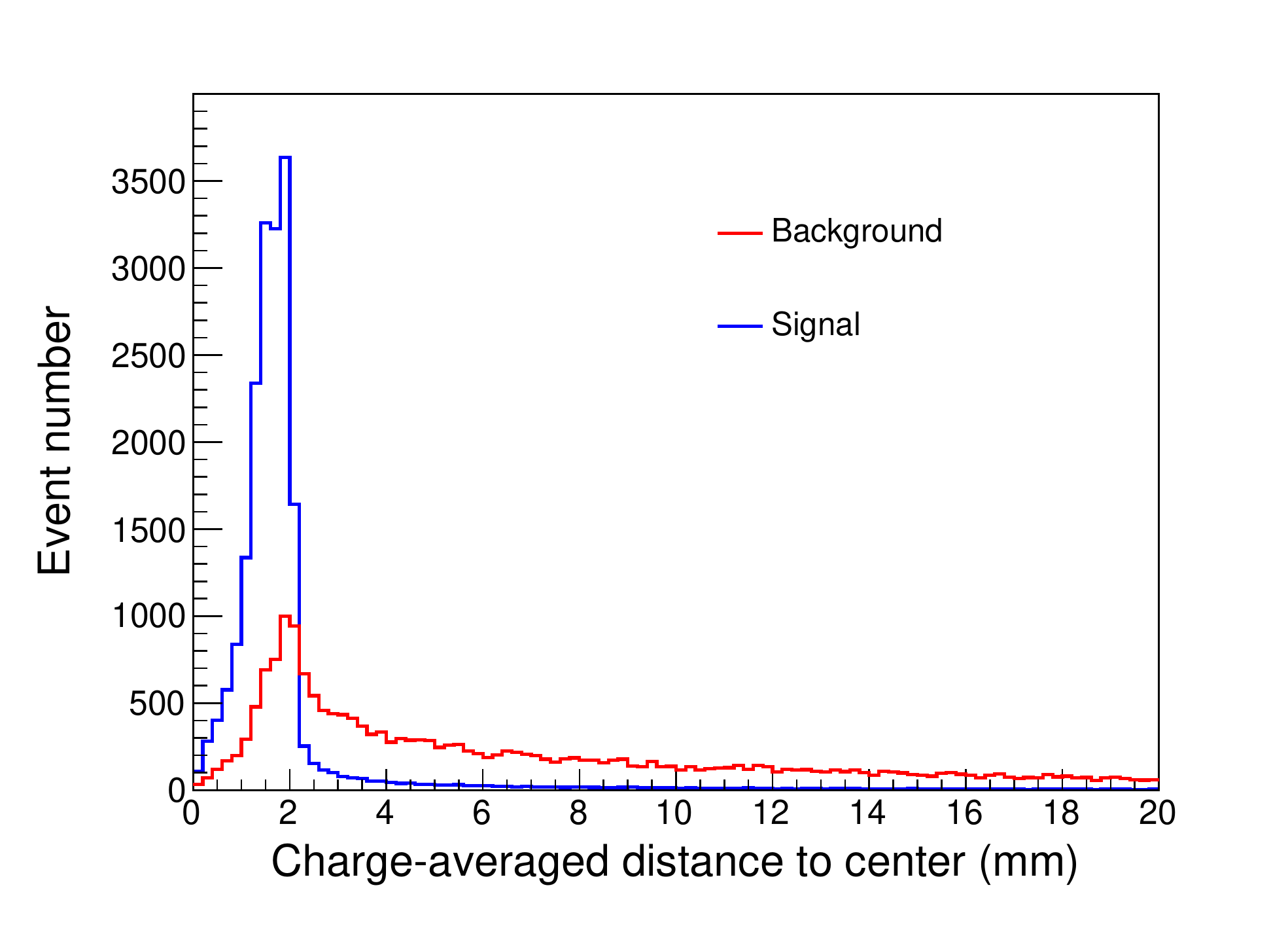}
    \caption{Charge-averaged distance to center in the $x$ coordinate for signal (blue) and background (red) events for a simulation with a 10~cm tile, 3~mm channel pitch, and 380~V/cm electric field. The same parameters are used in Figs.~\ref{fig:channel_num}--\ref{fig:sens_pitch}, except where stated otherwise.}
\label{fig:aved}
\end{center}
\end{figure}

\item[2.] ``Number of hit channels,'' $n_{chan}$.  For the same reason as described above, $\gamma$-ray events generally have more hit channels than $0\nu\beta\beta$ events. Fig.~\ref{fig:channel_num} shows the ``channel number'' variable, indicating the number of channels with a reconstructed collection signal above threshold, for signal and background events.

\begin{figure}[htbp]
\begin{center}
\includegraphics[width=0.5\textwidth]{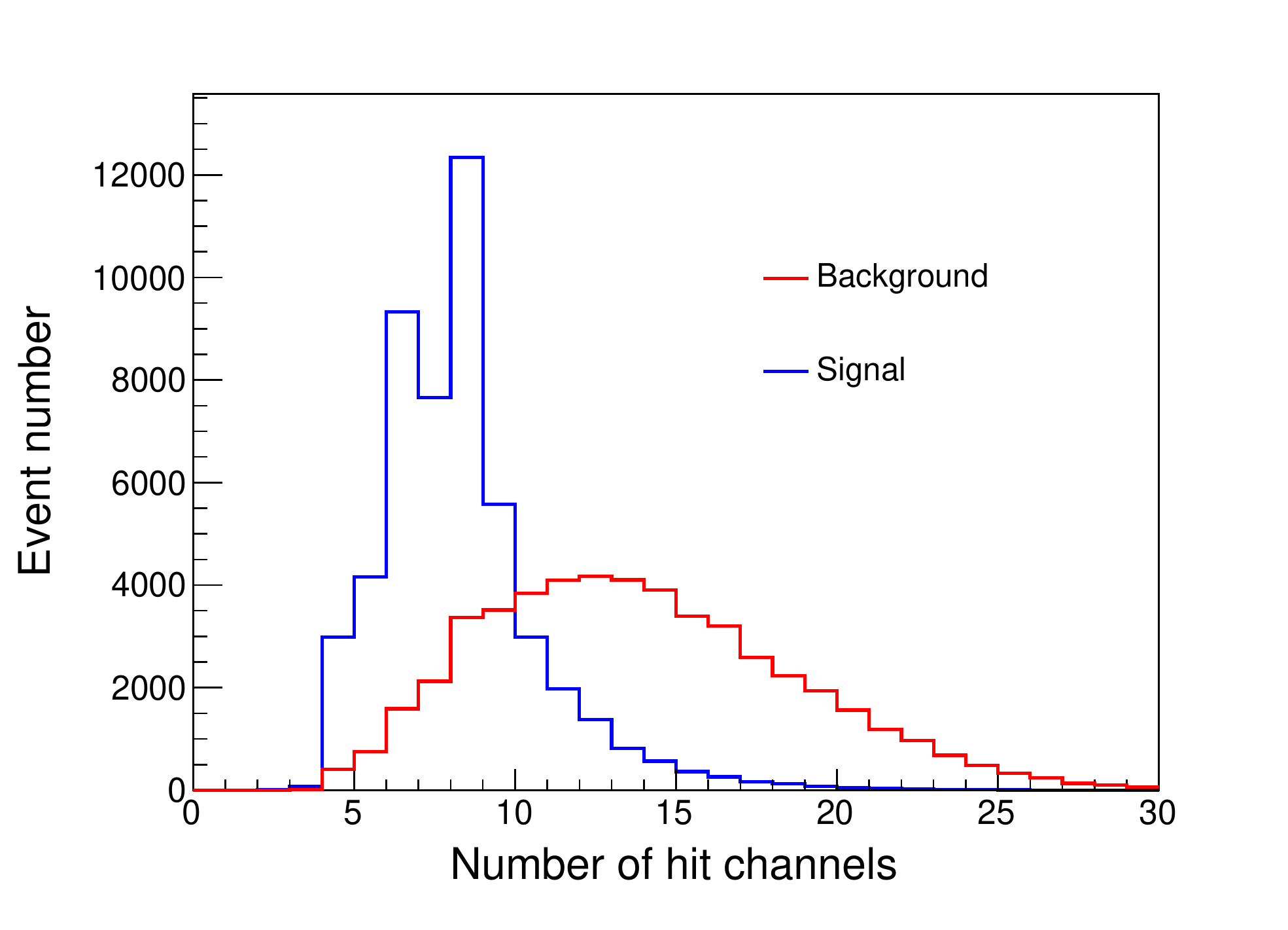}
\caption{Event ``channel number'' for signal (blue) and background (red) events. This variable has a sizable correlation with $d_x$ and $d_y$.  In order to check that such correlations do not lead to overtraining of the BDT, the distributions for both the training and testing samples are compared in Fig.~\ref{fig:bdtg}.}
\label{fig:channel_num}
\end{center}
\end{figure}

\item[3.] ``Maximum ratio of induction charge on a single channel in one event,'' $\textrm{Frac}_{ind}$. A peak search is performed on each charge waveform, and an ``induction charge'' is calculated as the difference between the maximum peak amplitude and the reconstructed charge. A non-zero induction charge is saved to the reconstructed event only when the reconstructed induction charge is greater than 3 times the channel RMS noise. A variable quantifying the maximum fraction of induction charge is defined as:
\begin{equation}
\textrm{Frac}_{ind} = \frac{q_{max}}{\sum q_m}
\label{eq:frac_ind}
\end{equation}
where $q_{max}$ is the maximum induction charge on a single channel, and $q_m$ is the induction charge on the $m^{th}$ channel in the event.  This variable approximately quantifies how sharply peaked the reconstructed induction signals are around the central channel.  This variable is expected to be more sharply peaked for signal events, where the charge is concentrated and produces more uniform induction signals on the central channel, than for background events that have a charge distribution that is more spread out. Fig.~\ref{fig:indfrac} shows the distribution of this variable for signal and background events.
\begin{figure}[htbp]
\begin{center}
\includegraphics[width=0.5\textwidth]{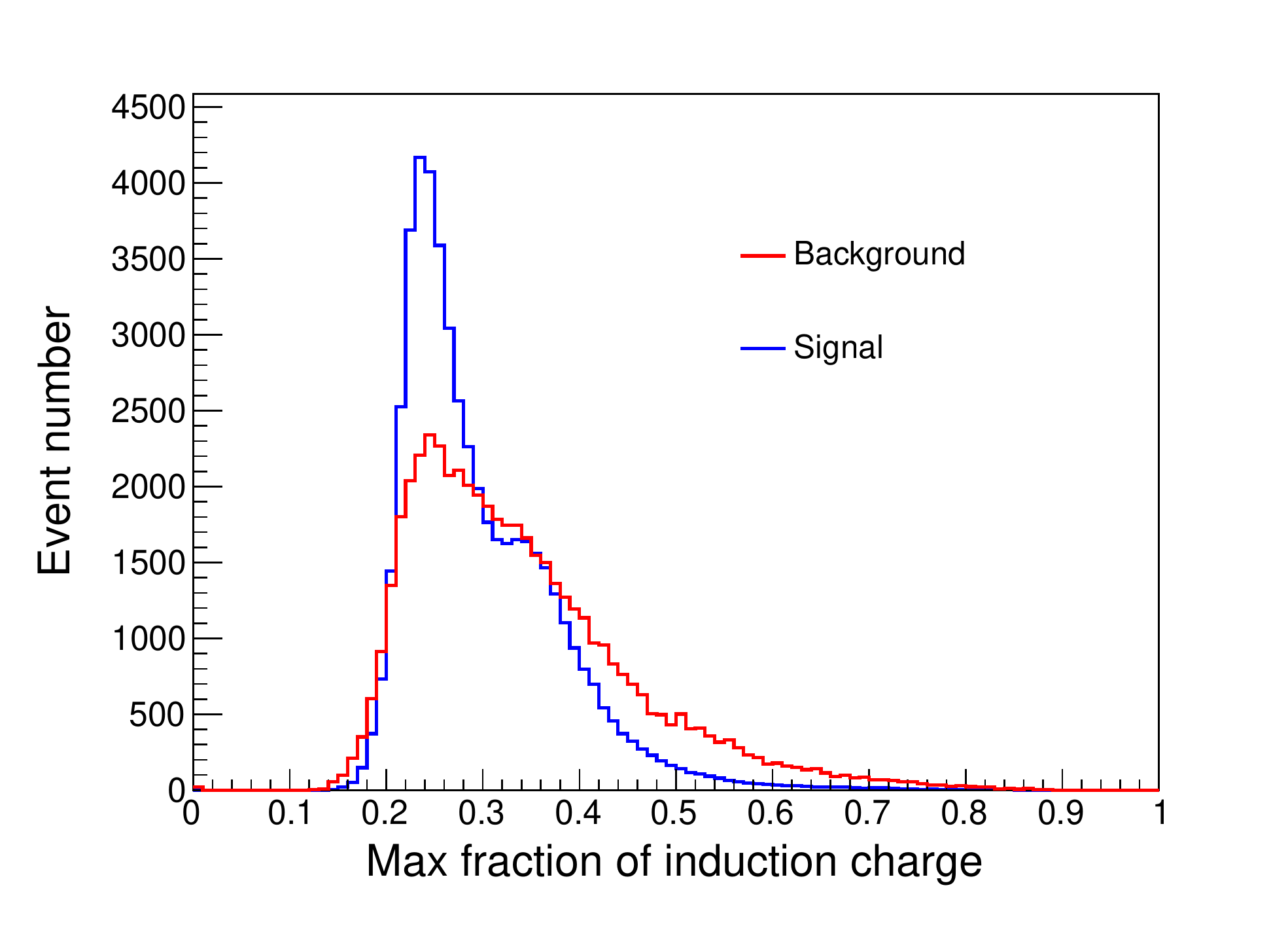}
\caption{Maximum fraction of induction charge in a single hit channel for signal (blue) and background (red) events.}
\label{fig:indfrac}
\end{center}
\end{figure}

\item[4.] ``Rise time of the summation of waveforms in an event,'' $t_{r}$. All the waveforms from channels with non-zero reconstructed charge energy are summed up to form a single waveform. A rise time is defined as the length of time between when the waveform first reaches 40$\%$ and 90$\%$ of its peak value. The starting amplitude threshold for the rise time is chosen to be sufficiently high to avoid the slow pre-pulse tail due to induced currents during the drift when the charges are far from the collection plane.  Because Compton scatters in $\gamma$-ray events typically have a larger distribution along the drift direction than signal events, they have a longer rise time distribution in this variable.  For the thresholds chosen, this variable is predominately sensitive to the time between when the upper and lower edge of the charge cloud are collected by the electrode. Fig.~\ref{fig:risingtime} shows the distribution of this variable in signal and background events.
\begin{figure}[htbp]
\begin{center}
\includegraphics[width=0.5\textwidth]{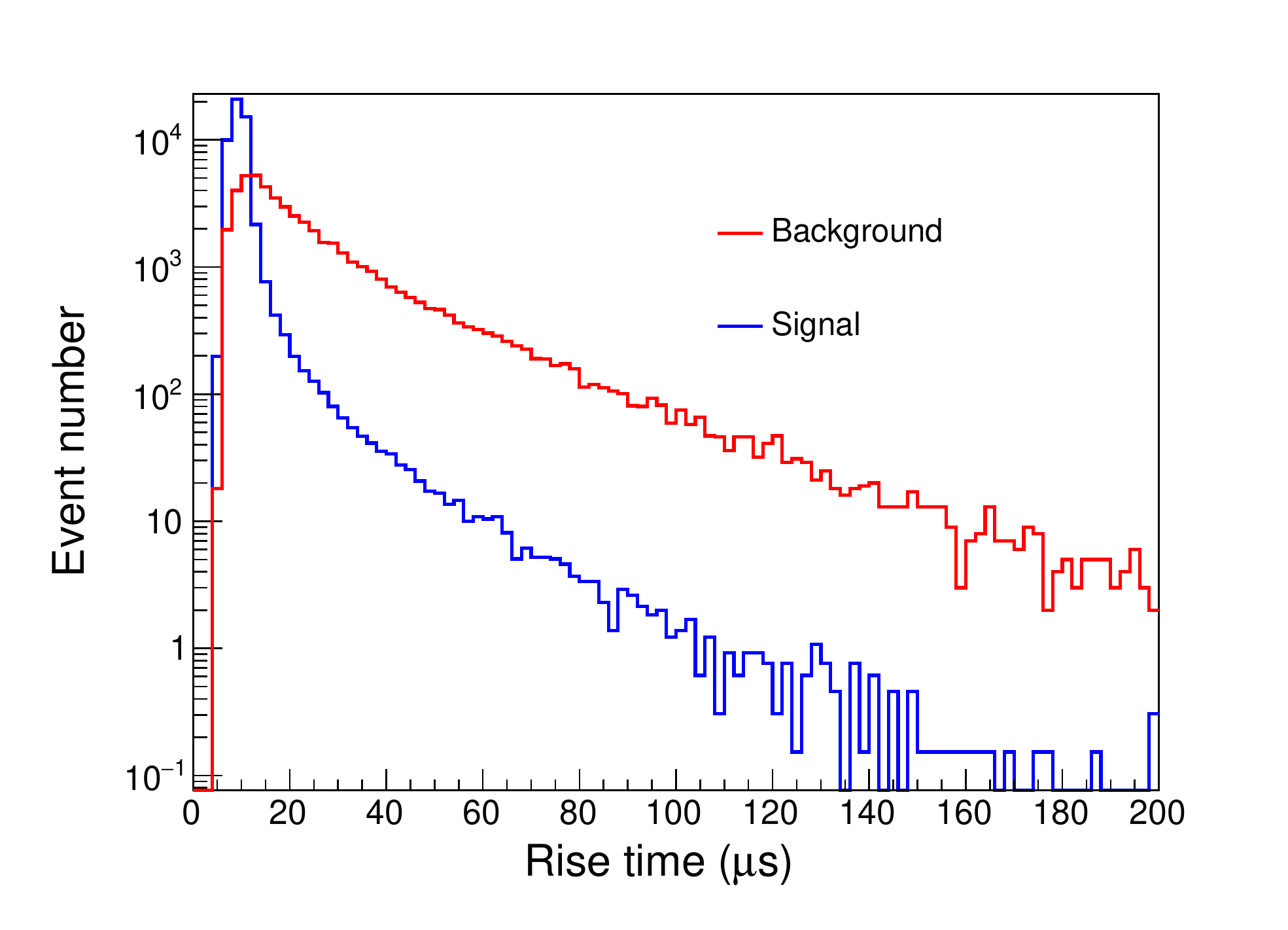}
\caption{Rise time distribution for signal (blue) and background (red) events. There is both a long tail in the rise time for background events (corresponding to well-separated Compton scatters), as well as a slight shift in the peak at small event size, due to closely spaced Compton scatters and the larger charge deposition of a single primary photoelectron (versus two lower energy electrons in $0\nu\beta\beta$).}
\label{fig:risingtime}
\end{center}
\end{figure}

\item[5.] Drift length, as described in Sec.~\ref{sec:recon_vars}. This variable shows differences in the distribution of external $\gamma$ events entering the detector, and is also used to correct for the effect of diffusion on the charge-averaged distance to center variable. The distribution of this variable is shown in Fig.~\ref{fig:drift_3mm}. 

\end{itemize}

\subsection{Comparison between detailed and fast charge simulation}
\label{sec:fast_charge_val}
\begin{figure}
  \centering\includegraphics[width=0.5\textwidth]{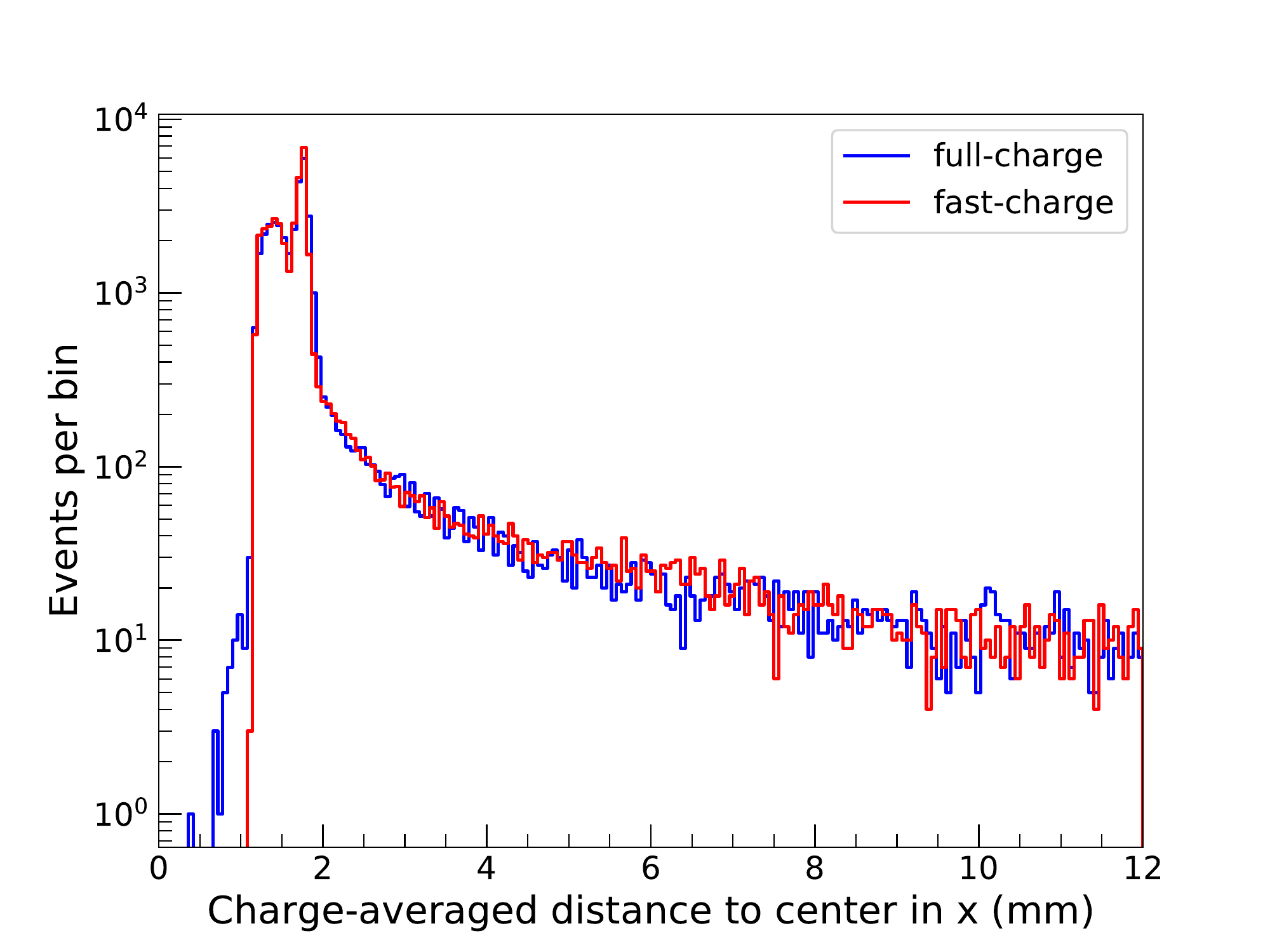}\includegraphics[width=0.5\textwidth]{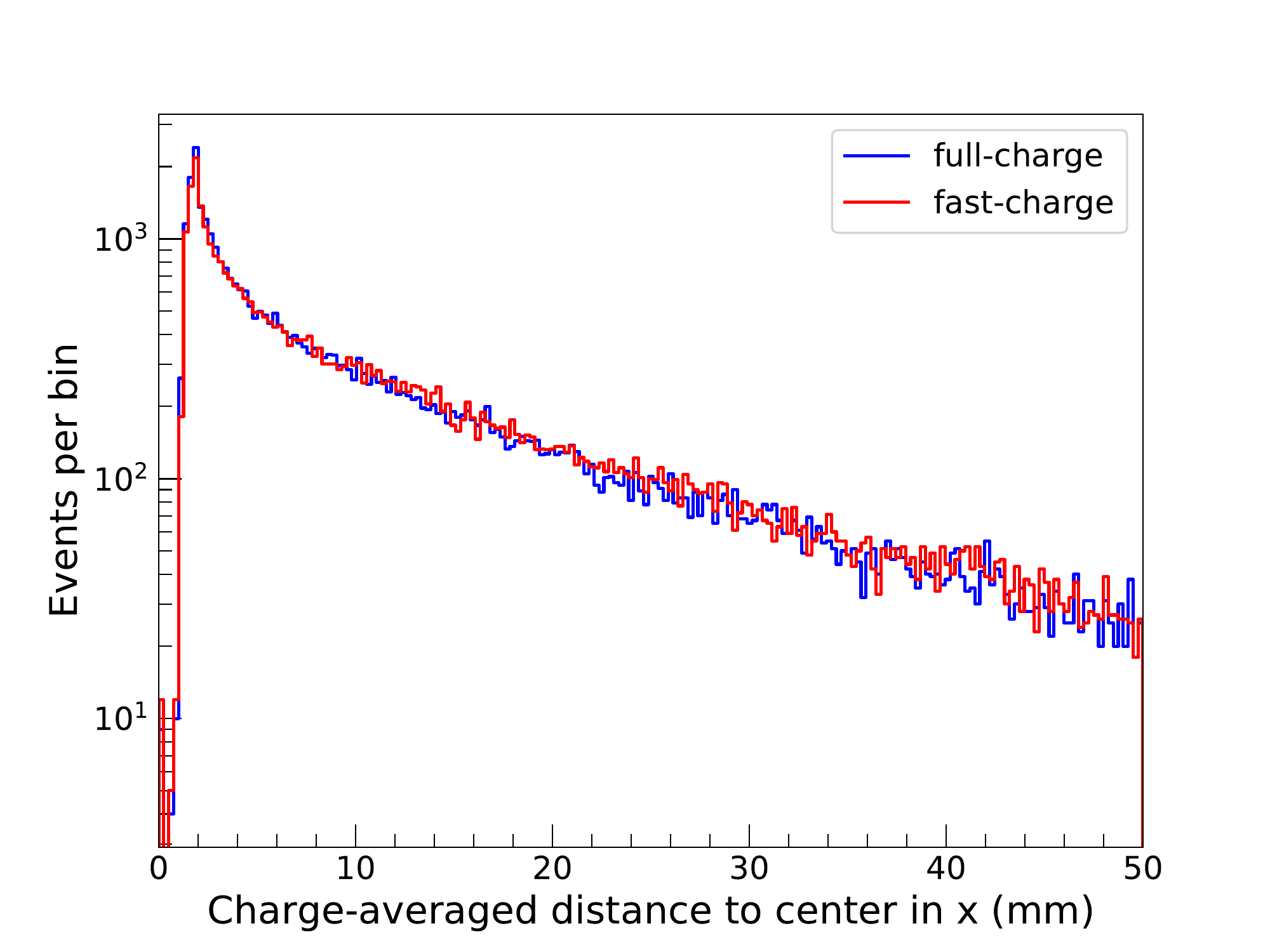}
  \caption[Comparison between full-charge sim.\ and fast-charge sim.]{Charge-averaged distance to center in $x$, compared between full charge simulation (blue) and fast charge simulation (red), for $\beta$ events (left) and $\gamma$ events (right).}\label{fig:simulation_comparison}
\end{figure}

\begin{figure}
  \centering\includegraphics[width=0.5\textwidth]{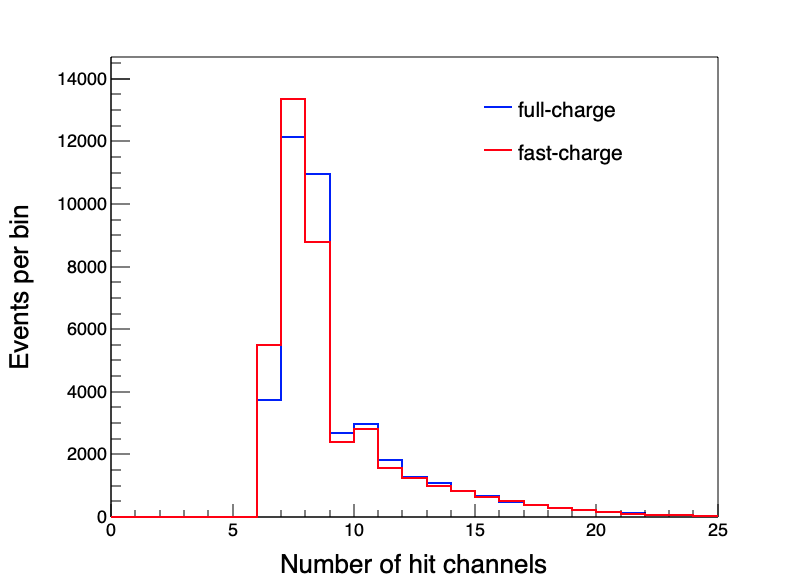}\includegraphics[width=0.5\textwidth]{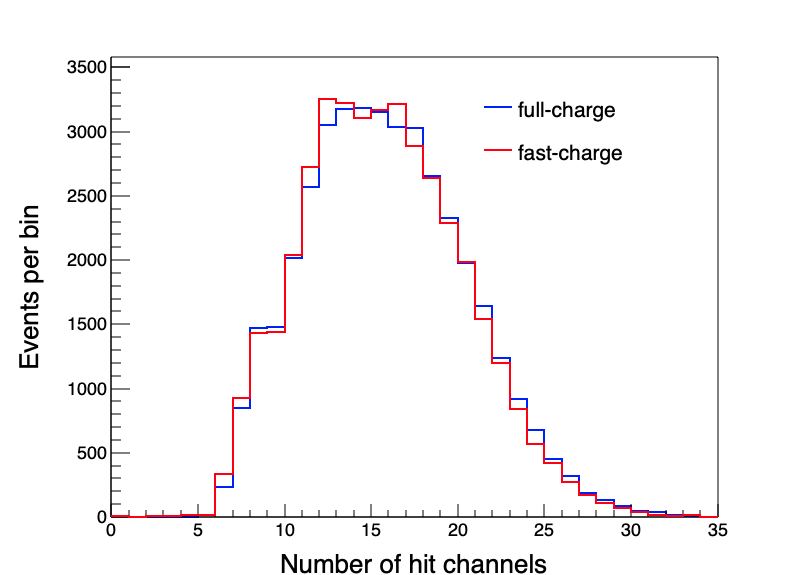}
  \caption[Comparison between full-charge sim.\ and fast-charge sim.]{Number of hit channels, compared between full charge simulation (blue) and fast charge simulation (red), for $\beta$ events (left) and $\gamma$ events (right).}\label{fig:nchannel_comparison}
\end{figure}

A comparison of the reconstructed topology variables from the full charge simulation and the fast charge simulation is shown in Figs.~\ref{fig:simulation_comparison}--\ref{fig:nchannel_comparison}. The primary electron and gamma events have energy of 2.5 MeV, and are produced on a plane of 30~cm radius that is parallel to the anode at the center of TPC. Both the fast charge simulation and detailed simulation use the same parameters for the drift velocity, diffusion, average noise, and energy threshold.

Since the fast charge simulation only produces energy deposits and not waveforms, it is not possible to directly compare results for all topology variables (e.g. $\textrm{Frac}_{ind}$ and $t_r$ can only be produced by the detailed charge simulation).  However, for the variables in which both simulations are able to produce distributions, good agreement is seen between the fast charge simulation and the full simulation.  While the fast charge simulation results are not used directly for the discrimination studies performed here, this agreement provides an independent cross-check on the detailed simulation and indicates that the fast charge simulation can provide distributions that agree with the detailed simulation for future studies making use of $d_x$, $d_y$, or $n_{chan}$.  


\subsection{Discrimination between signal and background using a multivariate discriminator}
\begin{figure}[!ht]
\begin{center}
\includegraphics[width=0.5\textwidth]{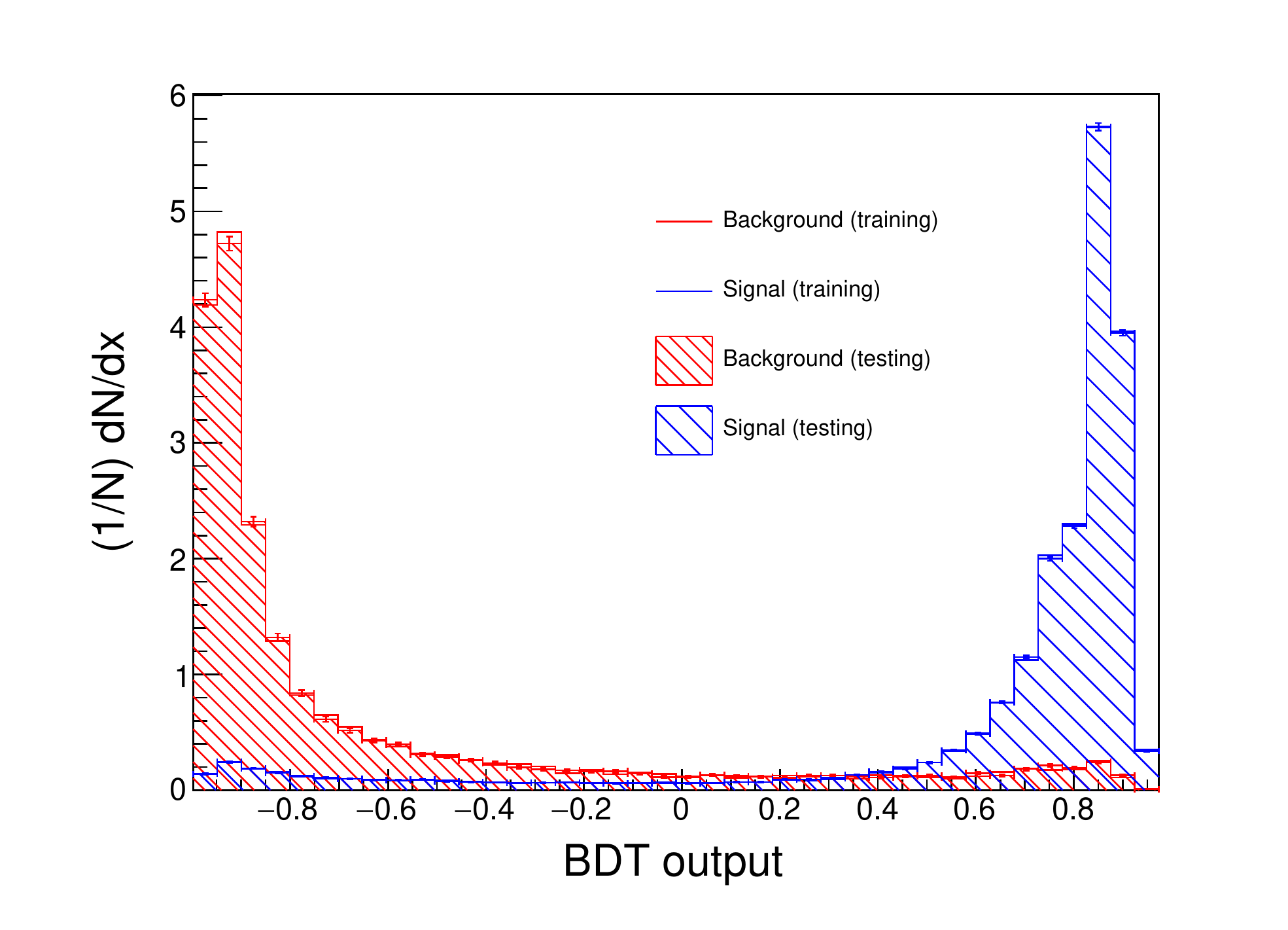}
\caption{Boosted decision tree output of signal (blue) and background events (red) in the training and testing data set. }
\label{fig:bdtg}
\end{center}
\end{figure}

The BDT is trained with $0\nu\beta\beta$ events as the ``signal'' event class, while the simulated events arising from U and Th activity in the nEXO MC geometry listed in Sec.~\ref{sec:g4_sim} are used as the ``background'' event class.  The trained discriminator is then used to identify events in a different set of MC simulation (the ``testing'' data set) that includes both $0\nu\beta\beta$ and background events. The output of an example BDT discriminator is shown in Fig.~\ref{fig:bdtg}. $0\nu\beta\beta$ events generally have a value close to 1, while background events have a value close to $-1$ in the BDT output.

\begin{figure}[b]
\begin{center}
\includegraphics[width=0.5\textwidth]{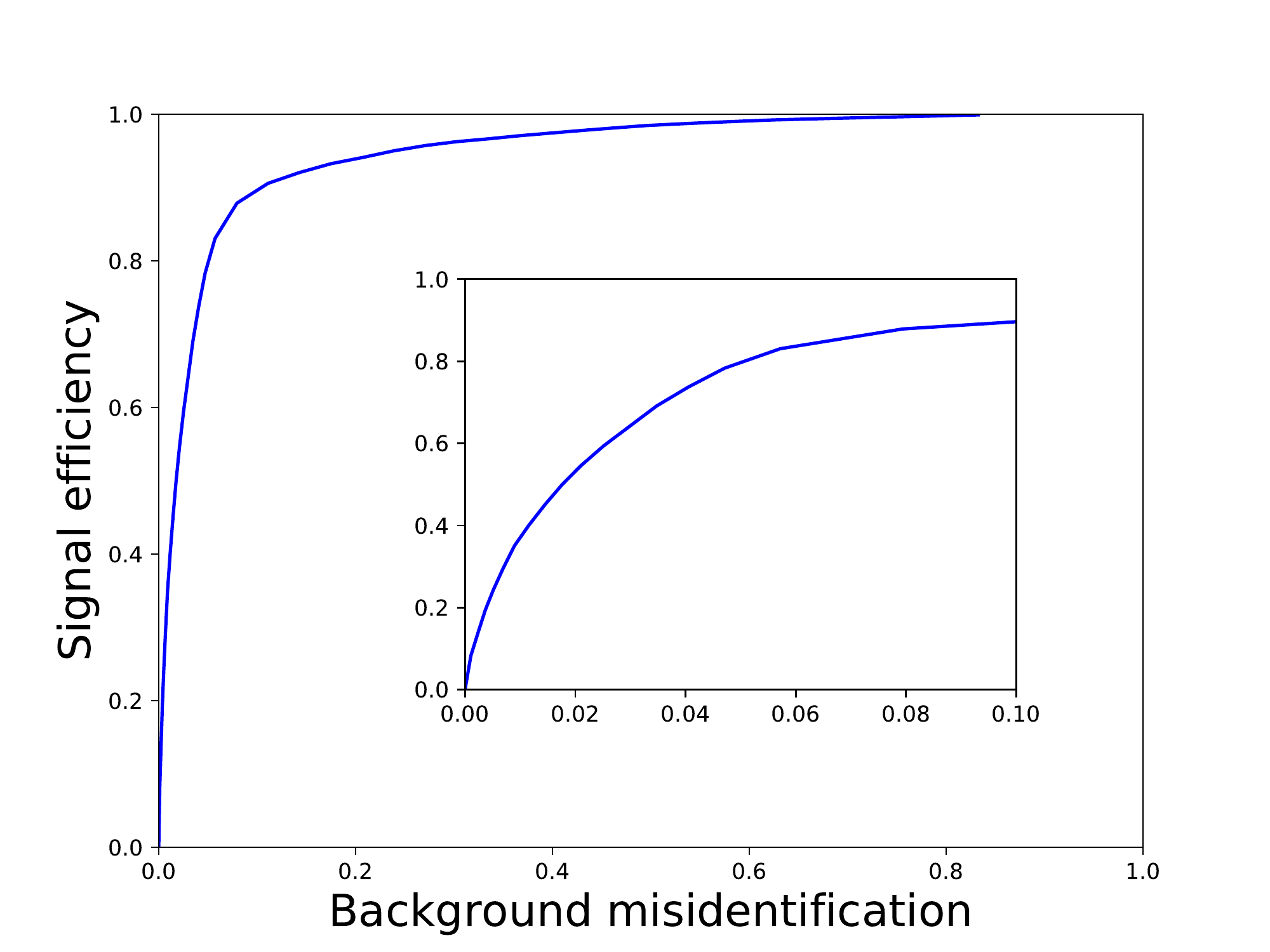}
    \caption{Signal efficiency versus background misidentification (ratio of background events classified as signal) for a varying cut on the BDT output distributions, as described in the text. The inset illustrates a zoomed in view of the curve at low values of background misidentification.}
\label{fig:roc_3mm}
\end{center}
\end{figure}

In EXO-200, the output of the BDT discriminator was used as an additional dimension in a multi-parameter fit~\cite{exo200_upgraded_0vbb}.  This allows the full distribution of the discriminator to be used in the analysis.  However, as an approximation to the sensitivity of the full fit, a simple 1D cut can be placed on the output of the BDT discriminator, and the signal efficiency and background discrimination as a function of cut position can be determined.  The resulting curve (known as the ``Receiver operating characteristic'' or ROC curve) 
corresponding to the BDT output distribution above is shown in Fig. \ref{fig:roc_3mm}. For each cut position, the vertical axis shows the signal efficiency versus the background misidentification.  

\subsection{Discrimination between signal and background using deep learning}
\label{sec:dnn}
Deep learning has been developing rapidly~\cite{lecun2015deep}, and is now widely used in computer vision, speech recognition, and for other tasks where classical algorithms are difficult to develop. Recently, deep learning has been introduced into particle physics as a promising tool to better extract information from data samples produced in particle physics experiments~\cite{Albert:2014awa, Aurisano:2016jvx, Delaquis:2018zqi}. In this section, a new algorithm for background rejection in nEXO is developed using deep learning, and its performance is compared to the multi-variate approach in Sec.~\ref{sec:bdt_discrim}. The
implementation of the algorithm is developed with an 18-layer deep residual network (ResNet-18)~\cite{he2016deep} using the PyTorch framework~\cite{paszke2017automatic}. 
\par The input to the network is produced with the charge simulation described in Sec. \ref{sec:chargesim}. Only channels with reconstructed charge greater than 3 times the channel noise are used to produce the data sample. 
The network requires an input of one or multiple two-dimensional (2D) arrays. To construct such 2D arrays, the coordinates of readout channels and current waveforms are used. The time sequence of the samples in each channel's current waveform are converted to a row in the input 2D array, with each row representing a different channel. The waveforms with 2~MHz sampling rate have a length of $\sim$1200 elements at most. However, the 2D array has a size limitation of 250$\times$250 samples due to
the memory limit of the GPUs used here. Therefore, the waveform is downsampled in time to fit the size limit. Fine sampling is adopted near the anode, and sparse sampling is used near the cathode (1~MHz for the last 40~$\mu$s of waveform, 500~kHz for the last 75 to 40~$\mu$s, 333~kHz for the last 75 to 100~$\mu$s, and 250~kHz for the rest of the waveform). This downsampling method preserves most of the information in the waveform as the charge approaches the anode, where the rate of change in the induced current is most rapid, while reducing the number of samples for the slowly changing current far from the anode. 

\begin{figure}[bp]
\begin{center}
\includegraphics[width=0.45\textwidth]{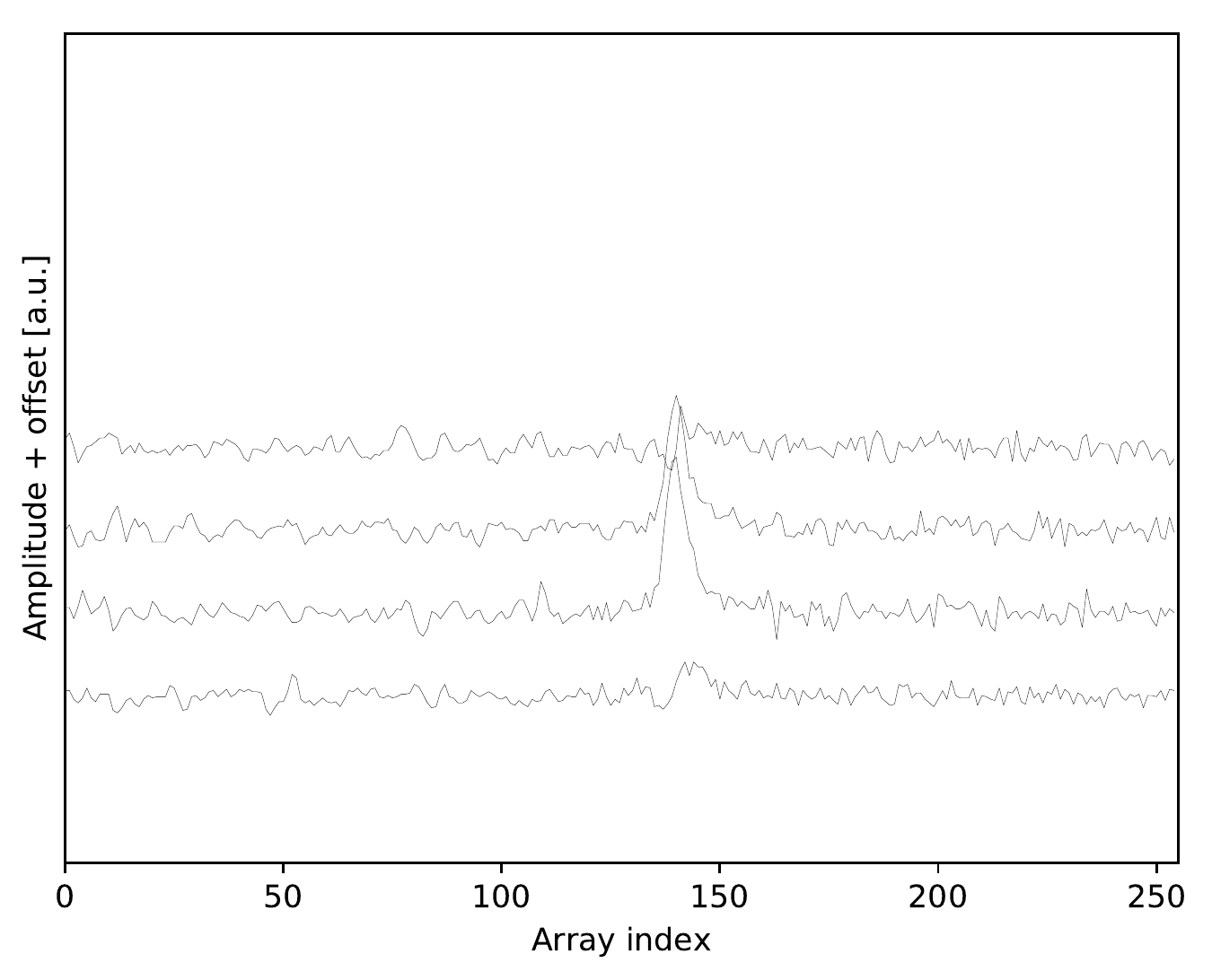}
~
\includegraphics[width=0.45\textwidth]{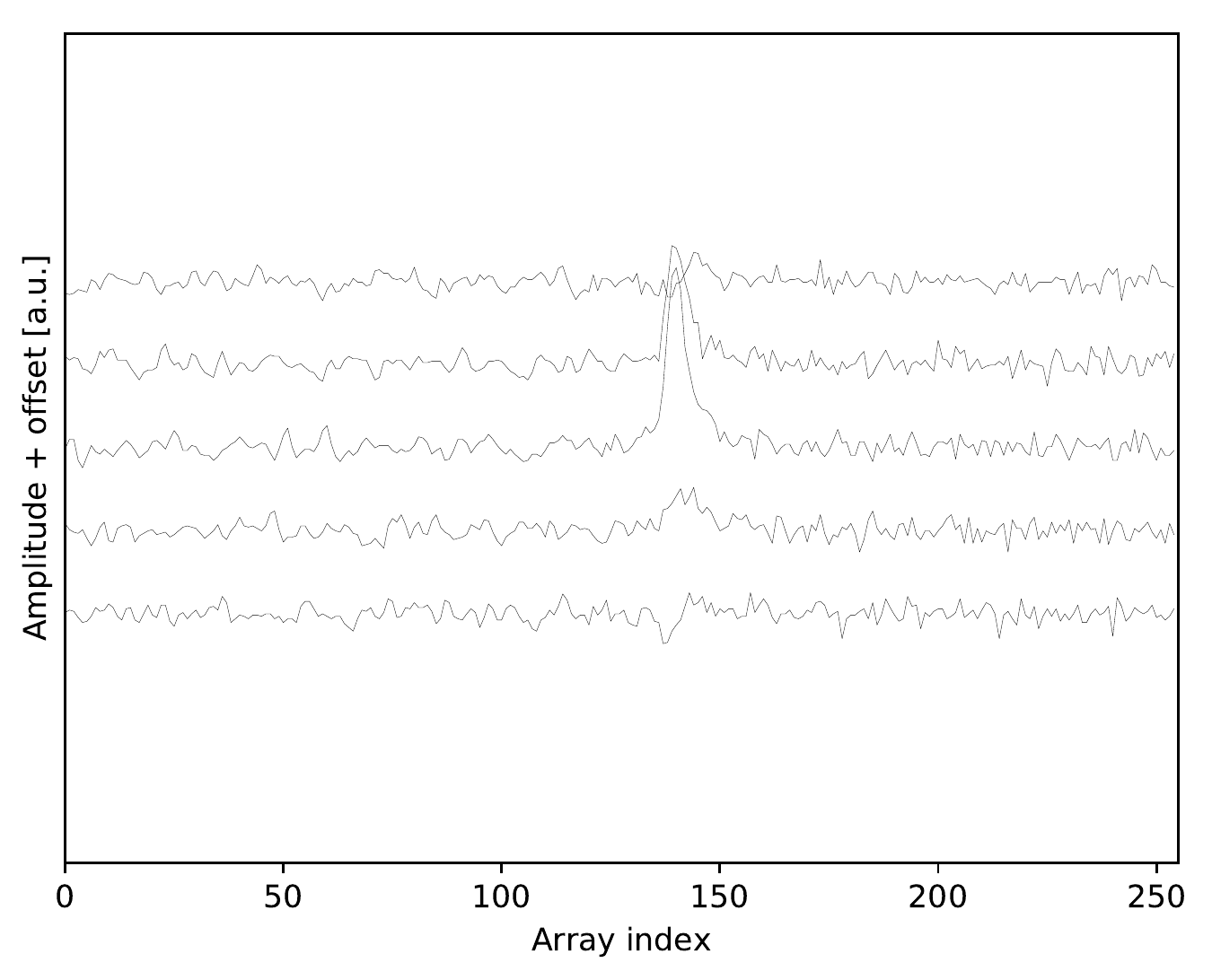}
    \caption{Examples of 2D arrays built with charge simulation waveforms in x-plane (left) and y-plane (right) for a simulated signal event. To zoom into the hit channels, the two plots use different vertical offsets. } 
\label{fig:dnn_input}
\end{center}
\end{figure}

Due to the architecture of the readout tile, each channel has a precise coordinate (e.g., in the direction of the 3~mm width of the readout strip) and an imprecise coordinate (e.g., in the direction of the 10~cm length of the strip). The channels are therefore split into two groups as two input channels to the network, and each group has the same precise coordinate. The precise coordinate value is then converted to row number. The row number has a span of $\sim$400 for a 3 mm pad size.  Since the whole anode needs to be cropped to fit within the maximum array size, only events with maximum size in the $x$ and $y$ coordinates less than 180$\times$(pad size) are used. By cropping the anode, about 3$\%$ of signal and background events are removed, although these are events with high channel multiplicity that are easy to identify and remove from the signal region using any discriminator. At the end, each element of the array is scaled to normalize the reconstructed charge of each event. Examples of these 2D arrays, which form the input to the network, are shown in Fig.~\ref{fig:dnn_input}.
\par
\begin{figure}[htbp]
\begin{center}
\includegraphics[width=0.5\textheight]{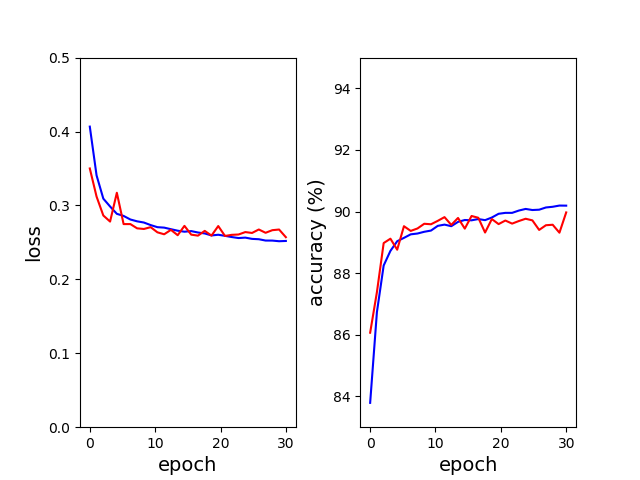}
    \caption{Loss and accuracy calculated for the training sample (blue) and validation sample (red) for 3~mm pad size as a function of the number of training iterations (epoch). } 
\label{fig:train_dnn}
\end{center}
\end{figure}

 60,000 simulated $0\nu\beta\beta$ decay and 50,721 background events are used to build the input arrays respectively. 80$\%$ of the events are randomly selected to train the DNN, and the other events are used as validation sample. The training of the network was carried out on one NVIDIA Tesla P100~\cite{nvidia2016p100}. The loss, a measure of difference between the DNN output and the truth information is computed, and a cross-entropy loss function is used in the training of the network. An accuracy variable, defined as the fraction of events with correct classification in the data set, is also computed. Fig.~\ref{fig:train_dnn} shows the loss and accuracy calculated for the training and testing samples as a function of the number of training iterations.

\begin{figure}[htbp]
\begin{center}
\includegraphics[width=0.35\textheight]{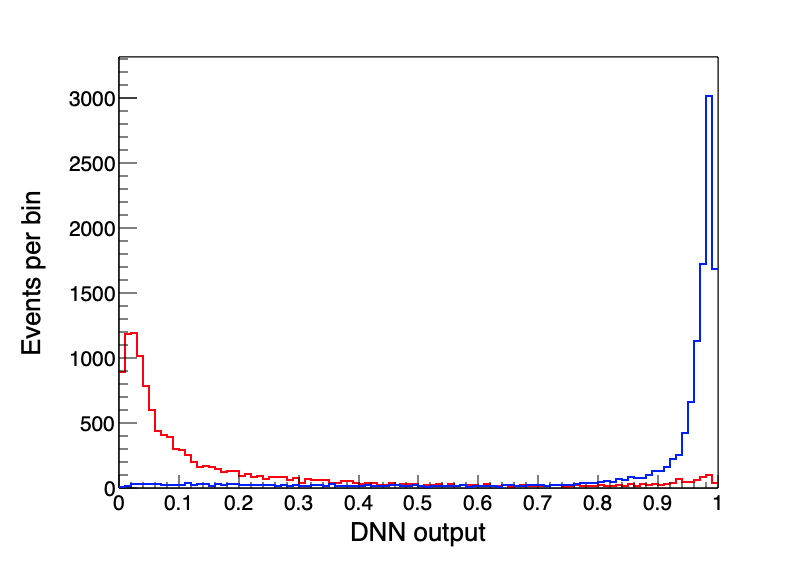}
    \caption{DNN output of the testing sample for signal (blue) and background (red) for  3~mm pad size. }
\label{fig:dnn_output}
\end{center}
\end{figure}

\begin{figure}[htbp]
\begin{center}
\includegraphics[width=0.35\textheight]{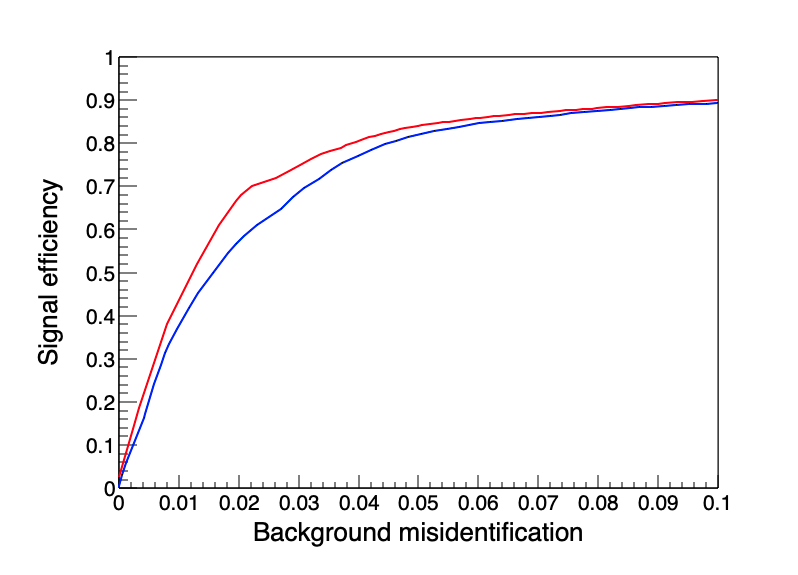}
    \caption{Comparison of the ROC curve built with the DNN output (red) and BDT output (blue) for 3~mm pad size. The horizontal axis shows only the range between 0--0.1, where the DNN method has a significant improvement over the BDT method.}
    \label{fig:dnn_roc}
\end{center}
\end{figure}

The output of the validation sample is shown in Fig.~\ref{fig:dnn_output}. Since the output of the DNN is normalized to one with the \textit{softmax} function~\cite{goodfellow2016deep}, it can be seen as the probability of the input event to be classified as signal or background. The output of the DNN for signal and background events is also converted to an ROC curve. The ROC curve built with DNN output is compared to that of the BDT method in Fig.~\ref{fig:dnn_roc}. The algorithm using deep learning demonstrates better
background rejection compared with the BDT method. For example, the simulation predicts that only 4$\%$ of background events survive in a signal sample with signal efficiency of 80$\%$ for 3~mm pad size. This performance can be compared to similar studies in EXO-200, where it was found that the discrimination power of the BDT could be improved to become comparable to the DNN with more advanced construction of input variables~\cite{Anton:2019wmi}. 

\section{Optimization of the detector design}
\label{sec:optimization}
\par A detailed simulation of the charge reconstruction in nEXO is important to optimize the design of the detector, especially the charge tile size, channel pitch, and electric field.  Since there are many tradeoffs involved in the detector optimization, it is important to accurately understand the impact of a proposed design change on the background discrimination for each set of detector parameters. This information can then be combined with engineering constraints and other requirements to determine the final charge tile and detector parameters~\cite{Kharusi:2018eqi}.
\subsection{Charge tile size and pitch}
A prototype charge tile has been produced with an edge length of 10~cm and a channel pitch of 3~mm, as shown in Fig.~\ref{fig:tileprototype}~\cite{Jewell:2017dzi}. In general, smaller channel pitches are expected to improve signal-to-background discrimination since they can allow the detailed structure of the charge deposits to be resolved.  However, smaller pitches require more readout channels, which would also increase the required number of corresponding cables and electronics channels, and increase
heat production from the readout electronics at the anode~\cite{Kharusi:2018eqi}. A larger number of cables and ASICs would also generally increase radioactivity.
Smaller pitches might also degrade the energy resolution since the charge is divided onto many channels, although the charge reconstruction algorithms used here indicate that the charge noise remains sub-dominant at all pitches considered.  The interplay of these effects on background discrimination is described below.

In combination with varying the channel pitch, using a charge tile with a longer edge could reduce the number of readout channels and tiles required for nEXO.  However, due to tile fabrication considerations, larger tiles would likely need to be formed by connecting four tiles with 10~cm edge length into a 2$\times$2 charge ``module,'' requiring the development of tile-to-tile connections~\cite{Kharusi:2018eqi}.  In addition, a larger charge tile increases the electronics noise per channel due to the larger capacitance of the longer strip. It also may increase the ambiguity of hits in the reconstruction, diminishing the background rejection efficiency. 

\begin{figure}[htbp]
\begin{center}
\includegraphics[width=0.5\textwidth]{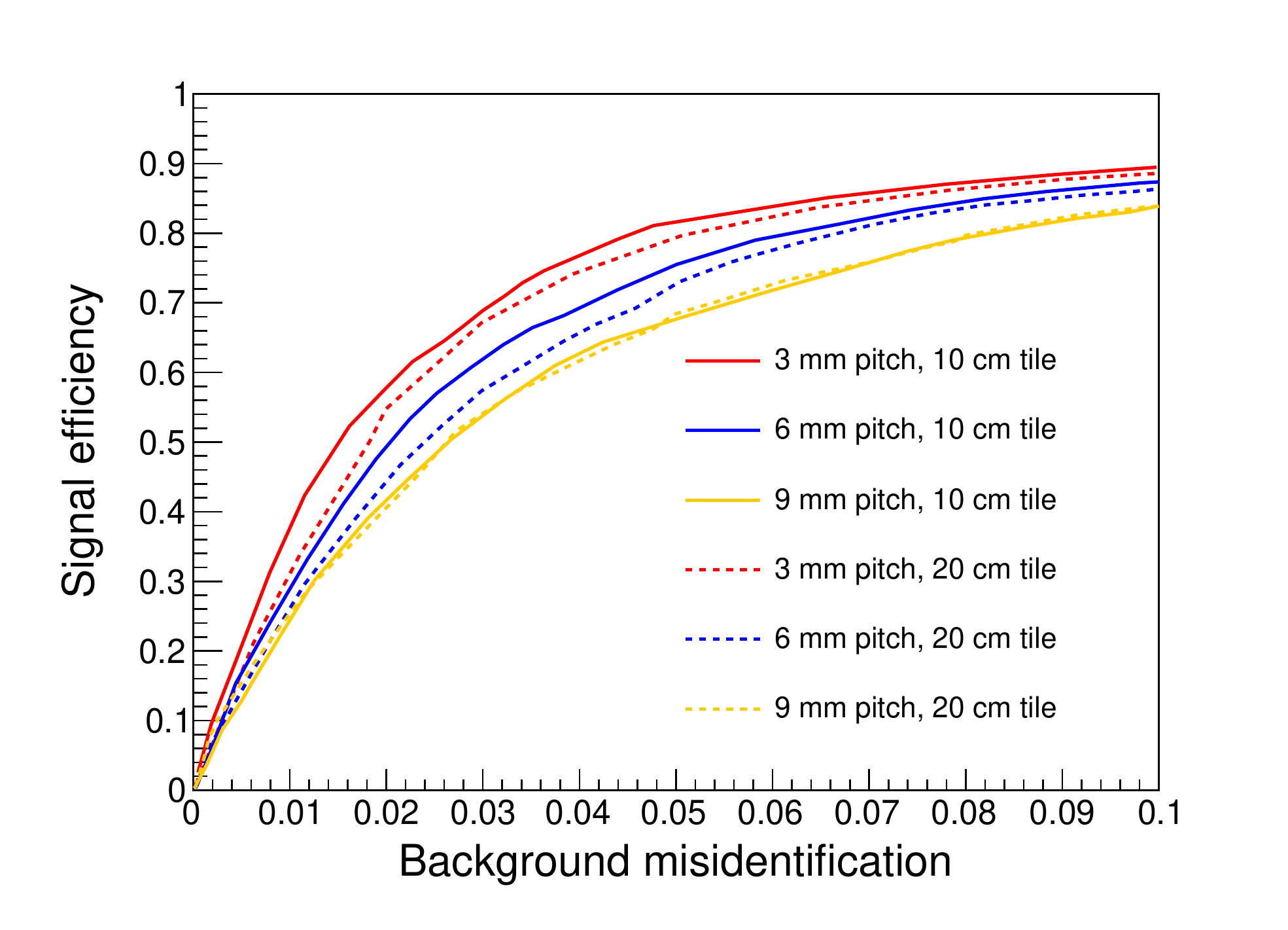}
    \caption{Comparison of the ROC curves with the BDT (the DNN method shows similar results) for different simulated module sizes and channel pitches.  The red/blue/yellow curves show the results for 3/6/9~mm pitch, with a single 10~cm length tile, while the red/blue/yellow dashed curves show the corresponding results for a 20~cm length charge module.} 
\label{fig:roc_pitch}
\end{center} 
\end{figure}

In order to evaluate the impact of pitch and tile size, we consider two tile sizes (10~cm and 20~cm), and three different channel pitches (3~mm, 6~mm, and 9~mm). Simulations of the charge drift, readout, and event reconstruction are performed for each combination of tile size and pitch, and the background rejection efficiency is compared. Fig. \ref{fig:roc_pitch} shows the ROC curves describing the background rejection achievable as a function of channel pitch and module size.  A smaller channel pitch leads to improvement in the simulated background rejection. The module size has a much smaller effect than the pitch, although a small improvement is seen for 10~cm size tiles compared to 20~cm tiles. The combination of 10~cm tile and 3~mm pitch has the best background rejection efficiency for the geometries considered.

\subsection{Electric field}
As described in Sec.~\ref{sec:chargesim}, both the electron drift velocity and diffusion depend on the electric field. A higher electric field results in an increase in the amount of energy collected as charge, leading to improved energy resolution. In addition, the shorter drift time at high electric field limits the effects of diffusion, leading to improved topological signal and background discrimination. However, due to the engineering challenges required for operating a large detector like nEXO at high electric fields~\cite{Kharusi:2018eqi}, the charge simulation has been used to quantify the expected change in sensitivity as a function of field. 

\begin{figure}[!htbp]
\begin{center}
\includegraphics[width=0.5\textwidth]{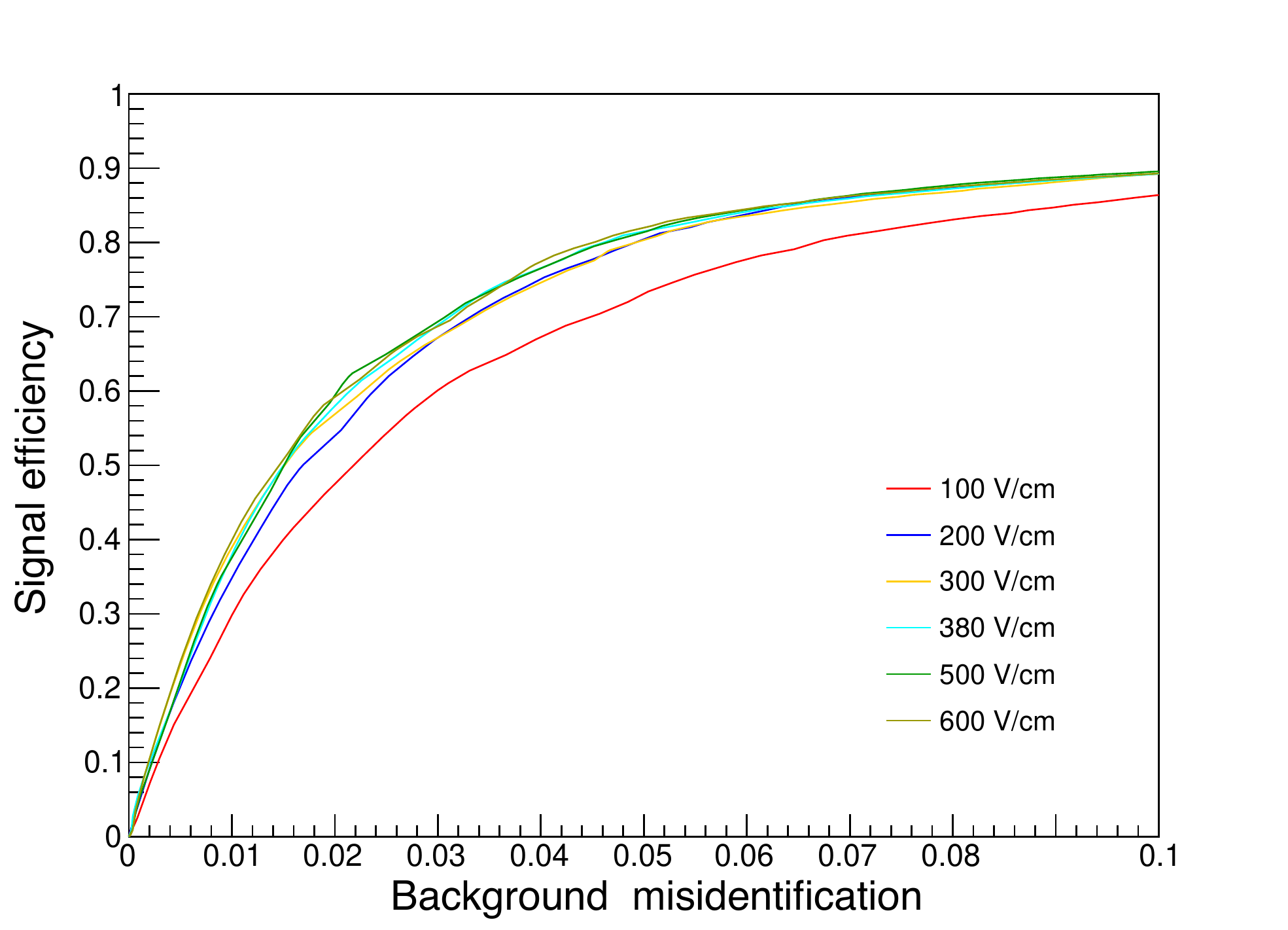}
\caption{Dependence of the ROC curves using the BDT on the electric field. }
\label{fig:roc_hv}

\end{center}
\end{figure}

To understand the effect of electric field and optimize it, a series of simulations are produced with electric fields from 100~V/cm to 600~V/cm (and a fixed tile geometry of 3~mm pitch and 10~cm length). The ROC curves giving the corresponding signal and background discrimination as a function of field are shown in Fig.~\ref{fig:roc_hv}.  While only a small improvement in the background rejection is observed going from fields of 200~V/cm to 600~V/cm, lower fields (e.g. 100 V/cm) do result in
reduced background rejection. The scaling in discrimination versus field results from the larger rate of change in the drift velocity for fields $\lesssim 200$~V/cm, which leads to additional smearing of the charge deposits due to diffusion during drift.  While these studies focus primarily on background discrimination, additional studies of the energy resolution and engineering complexity are required to determine the optimal operating field~\cite{hvreport}.

\section{Sensitivity estimation with charge-based signal and background discrimination}
\begin{figure}[htbp]
\begin{center}
\includegraphics[width=0.5\textwidth]{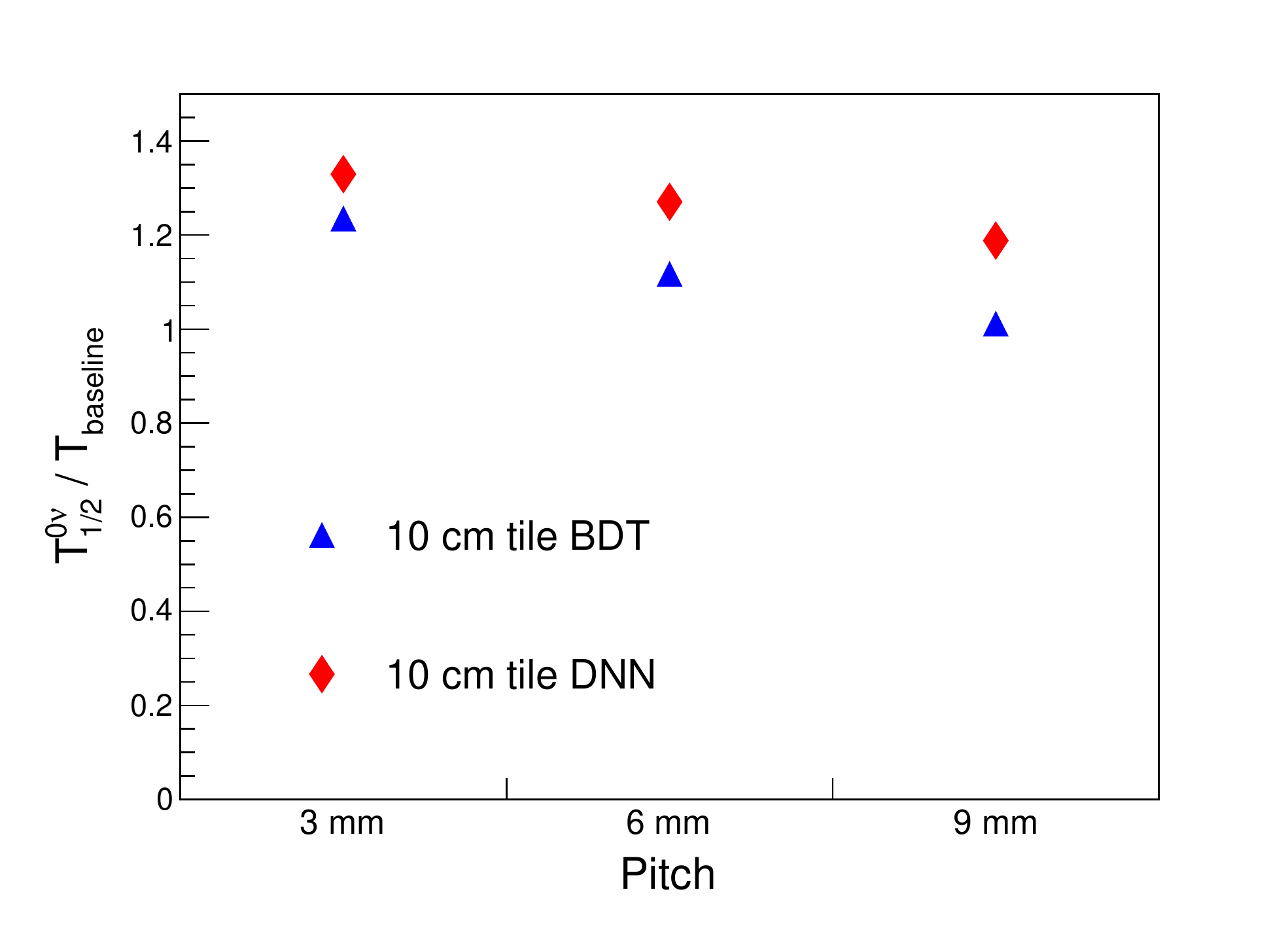}
    \caption{Relative change in the sensitivity to the $0\nu\beta\beta$ half-life in simulations with different charge tile pitch sizes.}
    \label{fig:sens_pitch}
\end{center}
\end{figure}

As shown in~\cite{Kharusi:2018eqi}, the sensitivity of nEXO to the $0\nu\beta\beta$ half-life is proportional to $B^{-0.35}$, where $B$ is the number of background events (for background levels in the central 3 tonnes of $\sim$0.9 cts/[FWHM tonne yr], as considered here). Background events are predominantly produced by $^{238}$U and $^{232}$Th within the energy range of interest for $0\nu\beta\beta$ decay events; therefore, $B$ is approximated by $^{238}$U and $^{232}$Th decays. To estimate the sensitivity with the BDT or DNN background rejection methods described above, the ROC curves are used to find the expected signal and background rate as a function of the cut position on the discriminator output.
Using the dependence of the sensitivity on the background rate above, the estimated $0\nu\beta\beta$ half-life sensitivity is then calculated as:
\begin{equation}
\frac{T^{0\nu}_{1/2}}{T_{baseline}}= \left(\frac{\epsilon_s}{\epsilon_s^{\mathrm{orig}}}\right) \left(\frac{\epsilon_b}{\epsilon_b^{\mathrm{orig}}}\right)^{-0.35} 
\end{equation}
where $\epsilon_s$ is the signal efficiency and $\epsilon_{b}$ is the fraction of background tagged as signal at a given point on the ROC curve.  The baseline half-life sensitivity, $T_{baseline} = 9.2\times 10^{27}$~yr, is based on the background discrimination values assumed in~\cite{Albert:2017hjq}.
The signal efficiency and background efficiency are used to scale the corresponding values assumed in~\cite{Albert:2017hjq}, $\epsilon_s^{\mathrm{orig}} = 0.85$ and $\epsilon_b^{\mathrm{orig}} =0.1$. By varying the position of the cut on the BDT or DNN output, the variation in the sensitivity can be determined from the fraction of background events tagged as signal and the estimated efficiency. At the optimal cut position, the sensitivity to the $0\nu\beta\beta$ decay half-life is found to improve by $\sim$20$~(32)\%$ relative to the sensitivity in~\cite{Kharusi:2018eqi} with the BDT~(DNN) method, for a simulation with 3~mm pad pitch and 10~cm tile edge. The best half-life sensitivity for $0\nu\beta\beta$ decay using simulations with different pad pitches are shown in Fig. \ref{fig:sens_pitch}. As shown in the figure, a smaller pad pitch results in better background rejection, although the simulations indicate that the tile size has a negligible effect on the background rejection.

\section{Conclusions}
A framework has been developed to simulate the electron drift and readout in LXe using the segmented charge tiles being designed for the nEXO experiment. A BDT method and a DNN method are developed to distinguish $0\nu\beta\beta$ events from background events based on the topology of the simulated charge signal. This BDT method reduces the background by 50$\%$ over the corresponding background rejection assumed in~\cite{Kharusi:2018eqi} with only 5$\%$ loss of signal efficiency. The DNN method further reduces the background by an additional 20$\%$ beyond the BDT method.  These results indicate that the ``aggressive'' goal for the overall nEXO background level in~\cite{Kharusi:2018eqi} could be reached through the use of the analysis techniques described here, with no changes to the assumed detector construction materials.  Such analysis techniques are independent of overall improvements to the radiopurity of the nEXO detector materials, and the plausible improvements from parallel work to reduce backgrounds through material screening and selection described in~\cite{Kharusi:2018eqi} would lead to further increases in the expected nEXO sensitivity.

Using this simulation framework, the background rejection ability of nEXO is determined for variations around the baseline detector design.  Among the three channel pitches of 3 mm, 6 mm, and 9 mm, a smaller pitch results in better background rejection for both the BDT and DNN methods. The difference in background rejection among the three pitches is slightly reduced using the DNN. The electric field has a small effect on background rejection when it is higher than
200~V/cm~\cite{hvreport}.

The nEXO sensitivity to the $0\nu\beta\beta$ decay half life is estimated with the background discriminators and simulation framework above. 
The best sensitivity 
found with these discriminators corresponds to a $\sim$20~(32)$\%$ improvement using a BDT (DNN) method relative to previous estimates~\cite{Albert:2017hjq}.

\section*{Acknowledgments}
This work has been supported by the Offices of Nuclear and High Energy Physics within DOE's Office of Science, and NSF in the United States, by NSERC, CFI, FRQNT, NRC, and the McDonald Institute (CFREF) in Canada, by IBS in Korea, by RFBR (18-02-00550) in Russia, and by CAS and NSFC in China. This work was supported in part by Laboratory Directed Research and Development (LDRD) programs at Brookhaven National Laboratory (BNL), Lawrence Livermore National Laboratory (LLNL), Oak Ridge National Laboratory (ORNL) and Pacific Northwest National Laboratory (PNNL).

\bibliographystyle{unsrt}

\bibliography{bibliography}
\end{document}